\documentclass[aps,twocolumn,amsmath,amssymb,floatfix,superscriptaddress,nofootinbib]{revtex4-1}

\usepackage{graphicx}
\usepackage[caption=false]{subfig}
\usepackage{dcolumn}
\usepackage{bm}
\usepackage{latexsym}
\usepackage{color}
\usepackage[colorlinks=true,linkcolor=blue,citecolor=blue]{hyperref}
\usepackage{xcolor}

\newcommand{\beq}{\begin{equation}}
\newcommand{\eeq}{\end{equation}}
\newcommand{\bali}{\begin{align}}
\newcommand{\eali}{\end{align}}
\newcommand{\kperp}{k_{\perp}}
\newcommand{\kpar}{k_{\parallel}}

\begin{document}

\title{Saturation of the asymmetric current filamentation instability under conditions relevant to relativistic shock precursors}
\author{Virginia Bresci}
\affiliation{Institut d'Astrophysique de Paris, CNRS -- Sorbonne Universit\'e, 98 bis boulevard Arago, F-75014 Paris}
\affiliation{CEA, DAM, DIF, F-91297 Arpajon, France}
\author{Laurent Gremillet}
\affiliation{CEA, DAM, DIF, F-91297 Arpajon, France}
\affiliation{Universit\'e Paris-Saclay, CEA, LMCE, 91680 Bruy\`eres-le-Ch\^atel, France}
\author{Martin Lemoine}
\affiliation{Institut d'Astrophysique de Paris, CNRS -- Sorbonne Universit\'e, 98 bis boulevard Arago, F-75014 Paris}

\date{\today}

\begin{abstract}  
The current filamentation instability, which generically arises in the counterstreaming of supersonic plasma flows, is known for its ability to convert the free energy associated with anisotropic momentum distributions into kinetic-scale magnetic fields. The saturation of this instability has been extensively studied in symmetric configurations where the interpenetrating plasmas share the same properties (velocity, density, temperature). In many physical settings, however, the most common configuration is that of asymmetric plasma flows. For instance, the precursor of relativistic collisionless shock waves involves a hot, dilute beam of accelerated particles reflected at the shock front and a cold, dense inflowing background plasma. To determine the appropriate criterion for saturation in this case, we have performed large-scale 2D particle-in-cell simulations of counterstreaming electron-positron pair and electron-ion plasmas. We show that, in interpenetrating pair plasmas, the relevant criterion is that of magnetic trapping as applied to the component (beam or plasma) that carries the larger inertia of the two; namely, the instability growth suddenly slows down once the quiver frequency of those particles equals or exceeds the instability growth rate. We present theoretical approximations for the saturation level. These findings remain valid for electron-ion plasmas provided that electrons and ions are close to equipartition in the plasma flow of larger inertia. Our results can be directly applied to the physics of relativistic, weakly magnetized shock waves, but they can also be generalized to other cases of study.
\end{abstract}

\maketitle

\section{Introduction}
The interpenetration of fast charged particle beams or plasmas gives rise to the current filamentation instability (CFI), often referred to as the Weibel instability~\cite{Weibel_1959, Fried_1959}. One of its remarkable features is to convert part of the free energy associated with the anisotropic momentum distribution into intense magnetic fields on skin-depth scales, even in the absence of pre-existing, coherent magnetization~\cite{Califano_1997,Bret+_2004,Bret+_2005,Bret+_2007,Bret_2008,Bret+_2010a,Bret_2010b,Wiersma+_2004,2007A&A...475....1A,Achterberg+_2007_II}. Its generic nature, its robustness and its physical implications have thus given it a compelling role in many fields of research, {\it e.g.}, high-energy density physics~\cite{Lee_1973,Molvig_1975,Silva_2002,Adam_2006,Debayle_2010}, laboratory astrophysics~\cite{Fiuza_2012,Huntington_2015,Lobet_2015,Warwick_2017,2019PhRvL.123e5002L,2020NatPh..16..916F}, cosmology~\cite{2001ApJ...563L..15G,2005PPCF...47A.205S}, and high-energy astrophysics, where it is thought to shape much of the nonthermal electromagnetic radiation from powerful explosive transients~\cite{Medvedev_1999,1999ApJ...511..852G,Silva_2003,Frederiksen+_2004,2010ApJ...710L..16K,2011A&ARv..19...42B,2011ApJ...737...55M,2013MNRAS.430.1280P,2013MNRAS.428..845L,2013ApJ...774...61K,2008ApJ...682L...5S,2008ApJ...673L..39S}.

Specifically, the CFI regulates the structure of weakly magnetized, collisionless shocks, which form through the counterstreaming of plasma shells at supersonic speeds~\cite{Moiseev_1963}. As such, it controls the production of high-energy particles and radiation in such environments, whether in the near~\cite{2015PhRvL.115l5001S} or the remote Universe~\cite{Medvedev_1999}. Because of their paramount consequences, those ``Weibel-mediated'' collisionless shocks have inspired a large body of literature, both in the subrelativistic~\cite{2008ApJ...681L..93K, 2010ApJ...721..828K, Bret+_2013, 2015Sci...347..974M, 2020ApJ...893....6B, 2020ApJ...904...12B, Ruyer_2015a, Ruyer_2015b, Ruyer_2016,Ruyer_2017} and relativistic regimes \cite{Lyubarsky_2006a, 2006ApJ...645L.129L, 2007ApJ...668..974K, 2008ApJ...682L...5S, 2009ApJ...695L.189M, Nishikawa+_2009, Keshet+_2009, Sironi+_2013, 2018MNRAS.477.5238P, 2019ApJ...886...54T, Lemoine+_2019a, Lemoine+_2019b, Lemoine+_2019c, Pelletier+_2019, 2020Galax...8...33V}.
Long-standing issues, with obvious phenomenological implications in the aforementioned domains of research, are the level of saturation of the CFI~\cite{Davidson_1972,Kato_2005,Califano_1998a,Okada_2007,Shvets+_2009,Cagas_2017,Grassi+_2017,Achterberg+_2007_II,2018ApJ...860L...1T,2019ApJ...877..137T} and the long-term evolution of the self-generated magnetic turbulence~\cite{2004ApJ...616.1065J,Medvedev_2005,Milosavljevic_2006a,2008ApJ...674..378C,2015JPlPh..81a4501L,Ruyer_2015a,Vanthieghem_2018}. Saturation is thought to occur through either transverse trapping of the particles in the magnetic filaments~\cite{Davidson_1972, Yang+_1994, Istomin_2011}, cyclotron gyration of the particles around the magnetic-field extrema~\cite{Moiseev_1963, Califano_1998a, Kato_2005, Istomin_2011}, or exhaustion of the available particle current~\cite{Kato_2005}. 
Most previous studies on the saturation of the CFI, and, to our knowledge, all those related to relativistic astrophysical systems, have considered symmetric configurations  in which the interpenetrating plasmas share similar characteristics (i.e., identical temperatures,  densities and drift velocities). In the precursor of collisionless shock waves, however, the interaction is strongly asymmetric, as it involves a hot, dilute beam of accelerated particles interacting with a cold, dense background plasma. 

The main objective of the present work is therefore to examine the saturation processes of the CFI in generic asymmetric configurations. In particular, we seek to determine which criterion holds, and whether this criterion applies to the beam, or to the background plasma, an ambiguity which obviously does not arise in symmetric configurations. We do so by confronting analytical predictions with particle-in-cell (PIC) simulations of initially unmagnetized, collisionless plasmas. For most of our study, we consider plasmas composed of equal mass species, interpenetrating each other at a relativistic velocity. This configuration is typical of the precursor of a relativistic shock propagating in a pair plasma, but it is also relevant for the study of the CFI in asymmetric electron-electron or ion-ion flows. We nonetheless extend our simulations to the case of electron-ion plasmas, in the ultrarelativistic and mildly relativistic regimes.

Our study is laid out as follows. In Section~\ref{sec:aCFI}, we recall the salient features of the CFI in asymmetric counterstreaming flows. In particular, we emphasize the notion of the preferred ``Weibel frame'', in which the instability is of a purely magnetic nature, and which becomes crucial in the asymmetric interaction regime. We then discuss the main saturation mechanisms and give the corresponding estimates of the maximum magnetic field energy. In Sec.~\ref{sec:PIC}, we present our PIC simulations for pair plasmas and analyze their results in light of the above mechanisms. We extend our analysis to the case of electron-ion counterstreaming configurations in Sec.~\ref{sec:epCFI}, and finally summarize our results and conclusions in Sec.~\ref{sec:conclusion}. 

\section{The asymmetric current filamentation instability}\label{sec:aCFI}

In this section as well as the next one, our initial setup comprises two counterstreaming, unmagnetized pair plasmas drifting along the $x-$axis. We note the beam with a subscript $_{\rm b}$ and the plasma with a subscript $_{\rm p}$. By convention, the beam corresponds to the population with the lower relativistic plasma frequency. The latter is defined as (cgs units are used throughout)
\begin{equation}
    \Omega_{\rm p \alpha}=\left(\frac{4\pi n_\alpha e^2}{\widetilde w_\alpha/ c^2}\right)^{1/2}\,,
    \label{eq:omegap} 
\end{equation}
where $e$ is the elementary charge, $c$ the velocity of light, $n_\alpha$ the proper number density and $\widetilde w_\alpha$ the enthalpy per particle of charged species $\alpha \in \{b+,b-, p+, p-\}$ in its initial state. Note that, in our notations, $n$ refers to a single charged species; it thus represents half of the initial total number density of the corresponding component (beam or plasma). Introducing the corresponding particle mass $m_\alpha$, adiabatic index $\widehat\Gamma_\alpha$ and proper temperature $T_\alpha$, one has $\widetilde w_\alpha = m_\alpha c^2 + \widehat \Gamma_\alpha k_{\rm B}T_\alpha/(\widehat \Gamma_\alpha -1)$. This implies $\widetilde w_\alpha \simeq m_\alpha c^2$ for a plasma of subrelativistic temperature ($k_{\rm B}T_\alpha/m_\alpha c^2 \ll 1$), and $\widetilde w_\alpha \simeq \widehat \Gamma_\alpha k_{\rm B}T_\alpha/(\widehat \Gamma_\alpha -1)$ for a relativistically hot plasma ($k_{\rm B}T_\alpha /m_\alpha c^2 \gg 1$). Given the inverse normalized temperature of species $\alpha$,  $\mu_\alpha \equiv m_\alpha c^2/(k_B T_\alpha)$, one has $\widetilde w_\alpha \simeq m_\alpha c^2$ and $\Omega_{\rm p \alpha}\simeq \omega_{\rm p \alpha}$ for a plasma of nonrelativistic temperature ($\mu_\alpha \gg 1$), but $\widetilde w_\alpha \simeq \widehat \Gamma_\alpha k_{\rm B}T_\alpha/(\widehat \Gamma_\alpha -1)$, and hence $\Omega_{\rm p \alpha}\simeq \omega_{\rm p \alpha}\sqrt{\mu_\alpha}/2$ (taking $\widehat \Gamma_\alpha = 4/3$) for a relativistically hot plasma ($\mu_\alpha \ll 1$).
In the following, use will also be made of
\begin{equation}
    \omega_{\rm p \alpha} = \left(4\pi n_\alpha e^2/m_\alpha\right)^{1/2} \,,
\end{equation}
the nonrelativistic plasma frequency of species $\alpha$. 

In this configuration, the counterstreaming instability can be described in wavenumber space $(k_{\parallel},\,\boldsymbol{k_\perp})$, in terms of the longitudinal $k_\parallel = \boldsymbol{k} \cdot \boldsymbol{\hat x}$ and perpendicular $\boldsymbol{\kperp}$ wavenumbers. This instability breaks into two main branches~\cite{Bret+_2010a}: purely transverse modes (the CFI) with $\kpar \ll \kperp$, and the so-called oblique two-stream modes, for which $\kpar \sim \kperp$. We neglect here the purely parallel electrostatic branch with $\kperp \ll \kpar$, which is usually subdominant in the relativistic limit. The oblique modes are essentially electrostatic, while the transverse CFI modes are essentially magnetic \cite{Bret+_2010b}, of direct interest to the present study. 

\subsection{The ``Weibel frame''}\label{sec: Wframe}

The notion of being ``magnetic'' or ``electrostatic'' is a frame-dependent statement, which must be made precise in a relativistic setting. This is discussed in detail in Ref.~\cite{Pelletier+_2019}, and we recap here the most salient features. The purely transverse CFI (meaning $k_\parallel \rightarrow 0$) develops through the pinching of the counterstreaming plasmas into filamentary structures oriented along $\boldsymbol{\hat x}$, each endowed with a net current. These structures are surrounded by toroidal magnetic fields $\boldsymbol{\delta B_\perp}$ and radial electric fields $\boldsymbol{\delta E_\perp}$. The CFI growth also comes with an inductive electric field component $\boldsymbol{\delta E_\parallel}$, oriented along the drift direction. In a first approximation, the latter field can be neglected, because its magnitude is of the order of $\vert \Im\omega/k_\perp c \vert \ll 1$ relative to the magnetic field component. The dominance of the magnetic component means $\delta B_\perp^2 - \delta E_\perp^2>0$ for each unstable wavenumber  $\mathbf{k_\perp}$, and hence that there exists a frame, moving at velocity (in units of $c$)
\begin{equation}
    \boldsymbol{\beta_{\rm w}} = \frac{\boldsymbol{\delta E_{\perp}} \times \boldsymbol{\delta B_\perp}}{\delta B_{\perp}^2}\,,
    \label{eq:vW}
\end{equation}
in which the transverse electric field component vanishes. In this frame, which we call the ``Weibel frame'', the CFI can be regarded as purely magnetic, up to the weak inductive component which cannot be erased by a Lorentz boost.

In the precursor of relativistic shocks, this frame gains special importance because the interaction between the beam of accelerated particles and the background plasma is so asymmetric that $\beta_{\rm w}\simeq 1$, meaning $\delta E_\perp\simeq\delta B_\perp$. It is crucial to properly characterize this frame, as it controls the heating and slowdown of the background plasma and because it greatly helps evaluate the scattering rate of suprathermal particles~\cite{Lemoine+_2019a}. Hence, the Weibel frame is connected to acceleration processes and has direct phenomenological consequences. 

Let us consider a set of initial beam $\left\{n_{\rm b},\,T_{\rm b},\,u_{\rm b \vert r} \right\}$ and plasma $\left\{n_{\rm p},\,T_{\rm p},\,u_{\rm p \vert r}\right\}$ parameters, where $u_{\rm \alpha \vert r} \equiv \gamma_{\rm\alpha \vert r} \beta_{\rm \alpha \vert r}$ denotes the $x-$component of the four-velocity of species $\alpha$, and $\beta_{\rm \alpha \vert r}$ and $\gamma_{\rm \alpha \vert r}$ are the associated normalized three-velocity (in units of $c$) and Lorentz factor, all defined in some reference frame (subscript $_{\vert \rm r}$). We use proper densities and temperatures unless explicitly specified otherwise. The corresponding Weibel frame velocity can be determined in the following two ways. 

In the linear phase of the CFI, one can define the Weibel frame as that in which the electrostatic component of the dispersion relation vanishes. This has been done in Ref.~\cite{Ruyer_2016} in the subrelativistic regime, and in Refs.~\cite{Pelletier+_2019,Lemoine+_2019a} in the relativistic regime. This amounts to setting the $\epsilon_{xy}$ component of the total dielectric tensor to zero, assuming $\boldsymbol{k_\perp} = k_\perp \boldsymbol{\hat y}$. This is not a trivial step, as the dielectric tensor itself depends on the solution to the dispersion relation, see  Ref.~\cite{Pelletier+_2019} for a discussion of the procedure.

Alternatively, one can describe the nonlinear phase of the instability as a quasistatic equilibrium between particles and fields, ordered along the transverse $y-$direction (a reduced 2D $x-y$ geometry is assumed throughout for simplicity) in a periodic sequence of current filaments. In a four-fluid (isothermal) description, the density of each component at equilibrium can be written as a function of the electromagnetic potentials, see Ref.~\cite{Vanthieghem_2018} for details. Setting the electrostatic contribution to zero imposes a relationship between the physical characteristics of the fluid, in the form
\begin{equation}
    \frac{n_{\rm b} \gamma_{\rm b \vert w}^2 \beta_{\rm b \vert w}}{T_{\rm b}} + \frac{n_{\rm p} \gamma_{\rm p \vert w}^2 \beta_{\rm p \vert w}}{T_{\rm p}}= 0 \,,
\label{eq:Wframe}
\end{equation}
where the normalized $x$-velocities $\beta_{\alpha \vert\rm w}$ and Lorentz factors $\gamma_{\alpha \vert\rm w}$ are here measured in the Weibel frame. The above equation can be solved to obtain the velocity of the Weibel frame in the reference frame. As it turns out, both methods give similar expressions for this velocity under conditions relevant to the precursor of relativistic shocks. Here, we rely on the latter method and make the result explicit, as follows.

Writing $\beta_{\alpha \vert \rm w}$ and $\gamma_{\alpha \vert\rm  w}$ in terms of $\beta_{\rm \alpha \vert r}$ and $\gamma_{\rm \alpha \vert r}$ through standard Lorentz transforms, one finds that the Weibel frame velocity, relative to the reference frame, can be expressed as 
\beq
    \beta_{\rm w \vert r}= \frac{Q_{\rm w} - \sqrt{Q_{\rm w}^2 -4}}{2} \, ,
    \label{eq:beta_w}
\eeq
where 
\beq 
    Q_{\rm w} = \frac{
    n_{\rm b} \gamma_{\rm b \vert r}^2 \left( 1+\beta_{\rm b \vert r}^2 \right)/T_{\rm b} + n_{\rm p} \gamma_{\rm p \vert r}^2 \left( 1+\beta_{\rm p \vert r}^2 \right)/T_{\rm p} }
    {n_{\rm b} \gamma_{\rm b \vert r}^2 \beta_{\rm b \vert r}/T_{\rm b} +  n_{\rm p} \gamma_{\rm p \vert r}^2 \beta_{\rm p \vert r}/T_{\rm p} } \,.
\eeq
The minus sign in Eq.~\eqref{eq:beta_w}  reflects the fact that $\beta_{\rm w \vert r} \simeq \beta_{\rm p \vert r}$ if the beam component becomes negligible: the turbulence is then mostly magnetic in the rest frame of the background plasma. Once $\beta_{\rm w \vert r}$ is known, the velocity of each species in a given reference frame can be Lorentz transformed to the Weibel frame.  

Henceforth, all velocities or Lorentz factors that do not carry a subscript $_{\vert \rm r}$ are understood to be defined in the Weibel frame. The  Weibel frame associated with the \emph{initial} state of the system is the reference frame in which our simulations will be conducted. Note that this frame can differ from the \emph{instantaneous} Weibel frame that results from the time-evolving properties of the beam and the plasma as the instability develops. 
This will be manifest in our simulations, and we will return to this point in Sec.~\ref{sec:PIC}.

\subsection{Linear stage of the CFI growth}
\label{sec: w growth}

Let us first consider the linear properties of the purely transverse CFI modes, that is, with wave vector $\mathbf{k_\perp} = k_\perp \boldsymbol{\hat y}$ and frequency $\omega \equiv i \Gamma_{\rm w}$, where $\Gamma_{\rm w}$ is the $k_\perp$-dependent growth rate. In the linear phase of the CFI, each mode grows as
\begin{equation}
    \delta B_z(k_\perp) = \delta B_0(k_\perp) e^{\Gamma_{\rm w}(k_\perp) t } \, ,
\end{equation}
where $\delta B_0(k_\perp)$ is the seed magnetic field fluctuation. Assuming that the magnetic spectrum ends up being dominated by modes of similar growth rate and seeded by comparable fluctuations, one can infer the instantaneous growth rate through
\begin{equation}
    \Gamma_{\rm w} =\frac{1}{2}\frac{{\rm d}}{{\rm d}t} \ln \left[ \frac{\langle \delta B_z(t)^2 \rangle}{\langle \delta B_z(0)^2 \rangle} \right]  \,.
    \label{eq:GR_linear_phase}
\end{equation}
The quantity in the rhs can be easily extracted from numerical simulations and directly compared with analytic estimations of $\Gamma_{\rm w}$. The latter involve rather heavy calculations of the dielectric tensor contained in the kinetic dispersion relation of which we will summarize here only the general key points.

The system is initially charge and current neutral with no equilibrium electromagnetic fields. 
Linearizing the Vlasov-Maxwell equations by expressing every perturbed physical quantity as $\delta \xi \propto e^{i( \boldsymbol{k}_\perp \cdot \boldsymbol{r}-\omega t)}$ yields the  dispersion relation of the CFI:
\begin{equation}
    \epsilon_{yy} \left(\epsilon_{xx}-1/\zeta^2 \right)=\epsilon_{xy}^2 \,,
\end{equation}
where $\zeta = \omega/k_\perp c$, and the elements of the dielectric tensor are given by
\begin{align}
    \epsilon_{ij} &= \delta_{ij}+ \sum_{\alpha} \frac{\gamma_{\alpha} \omega^2_{p \alpha}}{\zeta^2 k_\perp^2 c^2} \,\int \frac{u_i}{\gamma}\frac{\partial f_\alpha^{(0)}}{\partial u^j}\,d^3 u \\
    &+ \sum_\alpha \frac{\gamma_\alpha \omega^2_{p \alpha}}{\zeta^2 k_\perp^2 c^2}\,\int\frac{u_i u_j}{\gamma^2} \frac{\partial f_{\alpha}^{(0)}/ \partial u_y}{\zeta - \beta_y}\,d^3 u \,,
\end{align}
where $i,j=(1,2,3)$ and $\mathbf{u} = \gamma \boldsymbol{\beta}$.
Hence, the filamentation instability is generally not purely magnetic unless the off-diagonal term of the dielectric tensor vanishes, which would be the case for symmetric counterstreaming flows.

In the present work, we consider particle populations characterized by Maxwell-Jüttner momentum distribution functions.
As shown in Ref.~\cite{Pelletier+_2019},  approximate growth rates of the CFI can be obtained in two asymptotic limits that depend on the value of the parameter $\chi_\alpha = \gamma_\alpha \vert \zeta\vert / \sqrt{1-\zeta^2} $. For each plasma species, we define the \textit{hydrodynamic} limit in which the thermal velocity spread of the distribution function is, broadly speaking, smaller than the (imaginary) ``phase velocity'' of the waves, and the opposite \emph{kinetic} limit. More precisely, recalling that $\mu_\alpha = m_\alpha c^2/k_{\rm B} T_\alpha$, the hydrodynamic (resp. kinetic) limit for the cold plasma component corresponds to  $\tilde{\chi}_{\rm p} \equiv  \chi_{\rm p}  \sqrt{\mu_{\rm p} / 2} \gg 1$ (resp. $\ll 1$). For the relativistically hot beam component, the hydrodynamic (resp. kinetic) limit is rather defined as $\chi_{\rm b} \gg 1$ (resp. $\ll 1$), see ~\cite{Pelletier+_2019} for details.

We can thus derive two useful approximations of the maximum growth rate and associated wave number in terms of the nonrelativistic plasma frequencies of the plasma species, one in the fully kinetic regime -- meaning the kinetic approximation for both species -- and one in the combined hydrodynamical (beam) and kinetic (plasma) regimes, respectively,
\begin{align}
    \Gamma_{\rm w, k-k}  & \simeq \frac{(\omega_{\rm pb}^2 \mu_{\rm b})^{3/2} \gamma_{\rm b\vert p}^3 \beta_{\rm b\vert p}^3 }{ \sqrt{2 \pi \mu_{\rm p}}  \omega_{\rm pp}^2 + \frac{3 \pi}{2} \omega_{\rm pb}^2  \mu_{\rm b} \gamma_{\rm b\vert p}^3} \,, \nonumber \\
    k_{\rm \perp, k-k} & \simeq \sqrt{\frac{2}{3}\mu_{\rm b}}\gamma_{\rm b\vert p}\omega_{\rm pb} \,,  \label{eq:growth_rate_kk}
 \end{align}
 and
 \begin{align}
     \Gamma_{\rm w, k-h}  & \simeq \sqrt{\omega_{\rm pb}^2 \mu_{\rm b}} \,, \nonumber \\
    k_{\rm \perp, k-h} & \simeq (2\pi \omega_{\rm pb}^2 \mu_{\rm p} \mu_{\rm b})^{1/6}\omega_{\rm pp} ^{2/3} \,.
    \label{eq:growth_rate_kh}
\end{align}
The quantities $\beta_{\rm b\vert p}$ and $\gamma_{\rm b\vert p}$ represent the normalized three-velocity of the beam relative to the plasma and its corresponding Lorentz factor. It is important to stress that the above formulae have been derived solving the dispersion relation of the instability making the approximation of cold plasma ($k_{\rm B} T_{\rm p} \ll m_\alpha c^2$) and hot beam ($k_{\rm B} T_{\rm b} \gg m_\alpha c^2$) in the respective dielectric tensor. Furthermore, those approximations assume that the plasma moves at subrelativistic velocities with respect to the Weibel frame; consequently, it neglects terms of order $\mathcal O(\beta_{\rm p\vert w})$. Those formulas encompass the majority of the situations addressed in the following but not all; this will be made explicit. 

\subsection{Saturation criteria for the CFI}
\label{sec: Wsat}

In the early linear stage of the instability, the charged particles are deflected by magnetic field fluctuations with a polarity perpendicular to their initial drift velocity. As a result, particles of opposite charge from each component of the system (beam or plasma) are focused in different regions, forming transverse current modulations or ``current filaments''. Particles of opposite charges from both components concentrate in the same filaments where their currents add up; this amplifies the initial magnetic field perturbation, thus leading to the development of the instability. 

Eventually, the particle dynamics becomes modified by the fields so that saturation mechanisms take place. Ultimately, the CFI enters a strongly nonlinear stage, in which secondary instabilities, such as the merging of filaments of equal polarity, or the kink of current filaments, can arise, see Ref.~\cite{Vanthieghem_2018} for a detailed discussion. The transition between these two phases, i.e., saturation and the strongly nonlinear stage, is fraud with ambiguities, as filaments can coalesce while the current filaments keep building up through the CFI. We define here the saturation as the point at which the growth of the magnetic energy density is halted, or at least significantly reduced. This will be made clear in the figures that follow. 

To investigate the saturation of the instability, we will compare the temporal evolution of the magnetic field as extracted from simulations, with different criteria of saturation borrowed from the literature, which we summarize below. We emphasize a key difference with respect to the case of symmetric counterstreaming plasmas, which are more commonly envisaged.  In the asymmetric configuration, an ambiguity arises as to which component (beam or plasma) is eventually responsible for the saturation, and through which mechanism. For this reason, we discuss in the forthcoming paragraphs the saturation criteria as applied to a generic component. We will then apply each of them to the beam and to the plasma and compare those to the simulation results in the next Section.

\subsubsection{Transverse trapping}

The widely used trapping-based saturation criterion, first proposed by Davidson in the nonrelativistic regime~\cite{Davidson_1972}, and later generalized to the relativistic regime \cite{Yang+_1994, Lyubarsky_2006a, Achterberg+_2007_II, Kaang+_2009}, expresses the fact that, in the weakly nonlinear phase of the CFI, particles quiver transversely around the center of the filament (i.e., around a magnetic field node) in which they are focused. Assuming a harmonic $B$-field profile of amplitude $B$ and wavenumber $k_\perp$, a particle of Lorentz factor $\gamma$ and mass $m$ oscillates at the bounce frequency 
\begin{equation}
    \omega_{\rm B}=\left(\frac{e k_\perp \beta_\parallel B}{\gamma m}\right)^{1/2} \,.
    \label{eq:omegaB}
\end{equation}
The onset of saturation can be viewed as when the assumption of zero-order ballistic particle motion no longer holds. This occurs when 
$\omega_{\rm B}$ becomes comparable with the instability growth rate, $\Gamma_{\rm w}$.
Introducing $\langle \gamma \rangle$ the typical Lorentz factor of the considered species, the corresponding saturation magnetic field can thus be expressed as
\begin{equation}
    B_{\rm t} = \frac{\Gamma_{\rm w}^2}{k_\perp}\frac{\langle \gamma\rangle m}{\beta_\parallel e}\, .
\label{eq:trapp}
\end{equation}

\subsubsection{Magnetization limit}
\label{sss: magn. limit}

In the nonlinear phase of the CFI, the plasma can be modelled as an ensemble of cylindrical filaments of radius $r \simeq \lambda_\perp/4 \simeq \pi/2 k_\perp$, carrying a current density $j$. As the $B$-field grows in amplitude, the Larmor radius of the particles, $r_{\rm L} = \gamma \beta m c^2/e B$, shrinks, possibly up to the point where it becomes smaller than the filament radius. Particles then become \emph{spatially trapped} within the filaments in both the longitudinal and transverse directions, while orbiting around the $B$-field extrema. In the literature, this limit is often referred to as the ``Alfv\'en limit''~\cite{Kato_2005}. 
Similarly, particles gyrating at a Larmor frequency $\omega_L= eB/m \gamma$ higher than the instability growth rate can be regarded as \emph{temporally magnetized}. In either case, the linear approximation, which assumes rectilinear motion across the filaments, breaks down. The maximum value of the magnetic field set by this condition is then given by
\beq
    B_{\rm m} = \max{(B_{\mathrm{m},\,r_\mathrm{L}}, B_{\mathrm{m},\,\omega_\mathrm{L}})} \,, 
\label{eq: max B magn}
\eeq
where
\begin{equation}
    B_{\mathrm{m},\,r_\mathrm{L}}=  \frac{2}{\pi} k_{\perp} \langle \gamma \beta \rangle \frac{m c^2}{e}
    \label{eq: max_B_magn_sp}
\end{equation}
satisfies the spatial constraint and
\begin{equation}
    B_{\mathrm{m},\,\omega_\mathrm{L}} = \Gamma_{\rm w} \langle \gamma \rangle  \frac{m c}{e}
\end{equation}
the temporal one. Since the CFI is characterized by $\Gamma_{\rm w} \ll k_\perp c $ in relativistic shock precursors \cite{Pelletier+_2019}, it follows that usually $B_{\rm m} = B_{\mathrm{m},\,r_\mathrm{L}}$ if $\beta \sim 1$.
Similar saturation criteria were considered in \cite{Moiseev_1963, Medvedev_1999, Lyubarsky_2006a, Bret+_2013}. 

\subsubsection{Particle current limit}

The magnetic field is also bounded by above by the maximum current density that can sustain it \cite{Kato_2005}. This maximum current density corresponds to the current carried by one of the two oppositely charged species making up the component ({\it i.e.}, beam or plasma) under study. This limit thus tacitly assumes that, at maximum magnetic field, all the particles of a given component within a transverse length $\lambda_\perp/2 $ have undergone complete spatial separation in two adjacent filaments.
Assuming these have a uniform current density, the $B$-field created by a charged species of initial apparent density $\gamma n$ (with $\gamma$ characterizing here the drift motion) has a maximum strength
\begin{equation}
    B_p \simeq 2 \pi^2 \frac{e \gamma n}{k_\perp}  \langle \beta_\parallel \rangle \,.
    \label{eq: plimit}
\end{equation}

In the case of complete spatial separation, the contributions of counterstreaming species of opposite charge should add up within a filament. Yet in the asymmetric configurations addressed in the following, only the particle limit associated with the component that carries most, if not all of the particle current density, matters. We will therefore identify that component, which will turn out to be the beam component in most cases, and ignore the particle limit associated with the other (background plasma) component.

\subsection{Analytical estimates}\label{sec: sym case}

The hierarchy among the above saturation criteria depends on the characteristic wave number of the instability and the growth rate, given that
\begin{align}
    \frac{B_{\rm t}}{B_{\rm p}} & \sim \bigg(\frac{\Gamma_{\rm w}}{\omega_{\rm p}} \bigg)^2 \,, \\
    \frac{B_{\rm m}}{B_{\rm p}}  & \sim \left(\frac{k_\perp c}{\omega_{\rm p}}\right)^2 \,,
\end{align}
where $\omega_{\rm p}$ represents here the nonrelativistic plasma frequency of the component to which the saturation criterion is applied, and $k_\perp$ denotes the dominant transverse wavenumber. Considering first a cold symmetric counterstreaming configuration,  one has $\Gamma_{\rm w} \sim \omega_{\rm pp}$ and $k_\perp \gg \omega_{\rm pp}/c$ to leading order, \emph{e.g.} \cite{Wiersma+_2004, Bret_2008}. As a consequence, $B_{\rm t}\simeq B_{\rm p} \ll B_{\rm m}$, implying that the trapping and particle limits are equivalent and determine saturation. For symmetric counterstreaming hot plasmas, $\Gamma_{\rm w}$ is reduced to values below $\omega_{\rm pp}$, because it scales with the relativistic plasma frequency $\Omega_{\rm p} = \omega_{\rm p}\sqrt{\mu}/2$ and $\mu\ll1$. Consequently, the trapping criterion is expected to become more stringent than the other two. In addition, a relativistic temperature likely prevents the oppositely charged species of a given component from fully segregating from each other within a filament, further weakening the particle limit in this regime. 

In an asymmetric configuration, we identify distinct saturation limits for the beam and the plasma, using the respective superscripts $^{\rm b}$ and $^{\rm p}$. As a general trait of such configurations, we observe that the beam moves at relativistic velocities in the Weibel frame, while the drift of the background plasma is most often sub- or mildly relativistic. This can be read off Eq.~\eqref{eq:Wframe}, which relates the quantities $n_{\rm b}/T_{\rm b}\propto \Omega_{\rm pb}^2$ and $n_{\rm p}/T_{\rm p}\propto \omega_{\rm pp}^2 \mu_{\rm p}$. The beam is usually defined as the component with the smaller plasma frequency of the two, hence $\Omega_{\rm pb}\ll\omega_{\rm pp}$ suggests that $\gamma_{\rm b\vert\rm w}^2\vert\beta_{\rm b\vert w}\vert \gg \gamma_{\rm p\vert\rm w}^2\vert\beta_{\rm p\vert w}\vert$. Therefore, one must expect $B_{\rm t}^{\rm b} \gg B_{\rm t}^{\rm p}$. We anticipate that, for what concerns saturation through trapping,  only the larger of the two values $B_{\rm t}^{\rm b}$ and $B_{\rm t}^{\rm p}$ matters, and this trend will be confirmed by the simulations. 

We also expect, for the same reasons as above, that $B_{\rm t}^{\rm b} < B_{\rm p}^{\rm b}$ and $B_{\rm t}^{\rm b} \ll B_{\rm m}^{\rm b}$, because of the large temperature of the beam. Consequently, we may anticipate that the overall criterion for saturation will be set by the trapping limit of beam particles.

\section{PIC simulations results}\label{sec:PIC}

We have performed a number of 2D3V (2D in space, 3D in momentum) PIC simulations of counterstreaming electron-positron pair plasmas, which initially obey Maxwell-Jüttner distribution functions, using the massively parallel \textsc{calder} code \cite{Lefebvre_2003}. To resolve properly the initial Weibel instability, the cell size is set to $\Delta x =\Delta y=0.1 \, c/\omega_{\rm pp}$ and the simulations are run over $2 \times 10^4$ time steps of $\Delta t = 0.099 \, \omega_{\rm pp}^{-1}$ on a 2D $(x,y)$ grid of $2000 \times 2000$ cells. Henceforth, $\omega_{\rm pp}$ represents the nonrelativistic plasma frequency of each of the two charged species of the plasma component in its initial state, i.e., $\omega_{\rm pp}=\left(4\pi n_{\rm p}e^2/m_e\right)^{1/2}$ ($m_e$ is the electron mass). Each cell contains initially $100$ macro-particles per species, yielding a total number of about $10^9$ macro-particles. Time and length are normalized to the inverse nonrelativistic plasma frequency $\omega_{\rm pp}^{-1}$ and the plasma inertial length $c/\omega_{\rm pp}$. 

As previously mentioned, we aim to investigate the saturation of the current filament instability in an asymmetric interaction between a hot dilute beam and a cold, dense, inflowing plasma as it happens in the precursor of astrophysical collisionless shock waves in pair plasmas. For this reason, we initiate our study making use of initial parameters borrowed from a large-scale shock simulation corresponding to a relative upstream to downstream Lorentz factor of $10$, as described in \cite{Lemoine+_2019a}. The parameters of the beam and plasma populations, as measured in the downstream shock frame, are as follows (temperatures are given in units of $m_e c^2/k_{\rm B}$) : $\gamma_{\rm b\vert d} = 1.38$, $\gamma_{\rm p\vert d}= 9.67$, $T_{\rm b} = 45$, $T_{\rm p} = 0.2$, and $n_{\rm b}/n_{\rm p} = 0.1$. Those values are extracted from a region deep inside the precursor of the shock, where the background plasma has been slightly slowed down and heated to mildly relativistic temperatures. As announced, we then transform those initial parameters from the downstream shock frame to the Weibel frame. This gives the set of parameters indicated by (a) in Table~\ref{Tab:sim}, hereafter referred to as the reference run. Note that the plasma moves at subrelativistic velocities in this Weibel frame, while the beam is now ultrarelativistic. This difference demonstrates the importance of the Weibel frame regarding the development of the instability, and more importantly, regarding its saturation, since the saturation criteria depend on the inertia of the particles, which in turn depend on the reference frame.

The parameters of subsequent runs have been varied accordingly to fall in the region of the parameter space dominated by the CFI over electrostatic and oblique modes. In particular, we investigate a case where the initial beam proper density is tripled with respect to the reference case [run (b)], one in which the initial beam proper temperature is reduced by a factor of $1/3$ [run (c)], one with an initial beam Lorentz factor reduced by a factor of $1/3$ [run (d)]. Finally, we examine two more extreme configurations by reducing the initial temperature of the beam while increasing its initial Lorentz factor and the initial plasma temperature by a factor of $10$ each [run (e)] or $30$ each [run (f)]. The latter runs are of particular interest for the present study, because their parameters are such that the roles of background plasma and beam are interchanged with respect to other runs. 

What we refer to as the \textit{beam} is set in motion in the positive $x-$direction and represents the hot cloud reflected by the shock, which encounters the cold incoming \textit{plasma} streaming along the negative direction. Correspondingly, the transverse CFI generates an out-of-plane magnetic field component, $B_z$, aligned with the $\boldsymbol{\hat z}$ direction, and its associated electrostatic component $E_y$, along $\boldsymbol{\hat y}$. Since the simulation frame initially coincides with the Weibel frame, $E_y$ remains much smaller than $B_z$ during the initial development of the instability.
A stronger $E_y$ then emerges gradually, and as time progresses, the physical conditions of the plasma and/or the beam change, and so does the instantaneous Weibel frame. In particular, the filamentary structures start to move along $\boldsymbol{\hat x}$ at an approximately coherent velocity corresponding to the time-dependent value of $\beta_{\rm w}$. To discriminate between the various saturation criteria, the magnetic field is directly extracted from PIC simulations and compared with the theoretical estimates of the saturated $B$-field given in Sec.~\ref{sec: Wsat}.

In what follows, we focus on the linear and saturation phases of the instability, while the late-time evolution is left aside and treated in Sec.~\ref{subsec: late time ev}.

\begin{table}
    \begin{tabular}{ |p{0.6cm}|p{1cm}|p{1cm}|p{1.2cm}|p{1.2cm}|p{1.8cm}|} 
    \hline
 Run                                  	              &   $\gamma_{\rm b}$ &$\gamma_{\rm p}$ &$ T_{\rm b}/ \gamma_{\rm b}$ & $ T_{\rm p}/\gamma_{\rm p}$  & $\gamma_{\rm b} n_{\rm b} /\gamma_{\rm p} n_{\rm p}$ \\ 
    \hline
 (a)                       	               & $18.9$            &  $1.01$      &            $2.4$                &           $0.2$                &     $1.9$ \\ 
 (b)      & $16.7$            &  $1.05  $      &            $2.9$                &           $0.2$                &     $4.6$     \\ 
 (c)              	      & $16.7$            & $1.05  $       &            $0.95$                &          $0.2$                 &     $1.5$   \\ 
 (d)       & $40.8  $            & $1.2  $       &            $1.09$                &          $0.2$                 &     $3.5$   \\ 
 (e)     & $25. $          & $ 5.3 $     &  $0.2$          &          $0.4$                 &     $0.5$    \\ 
 (f)   & $25.  $          & $ 15.4 $   &  $0.06$        &          $0.4$                 &     $0.2$   \\ 
    \hline
    \end{tabular}
    \caption{\label{Tab:sim} Summary of simulation parameters for pair plasmas. Run (a) is the reference simulation. The parameters of the other runs differ from those of run (a) as follows: (b) $n_{\rm b} \times 3$; (c) $T_{\rm b}/3$; (d) $\gamma_{\rm b\vert d} \times 3$; (e) $\gamma_{\rm b\vert d}/3$,(e) $\gamma_{\rm b\vert d} \times 10$, $T_{\rm p} \times 10$, $T_{\rm b}/10$; (f) $\gamma_{\rm b\vert d}\times 30$, $T_{\rm p}\times 30$, $T_{\rm b}/30$. The table gives the simulation parameters once transformed to the Weibel frame. Temperatures are given in units of $m_e c^2/k_{\rm B}$.}
\end{table}


\subsection{Reference run}\label{sec:ref PIC}

\begin{figure}
        \includegraphics[width=0.42\textwidth]{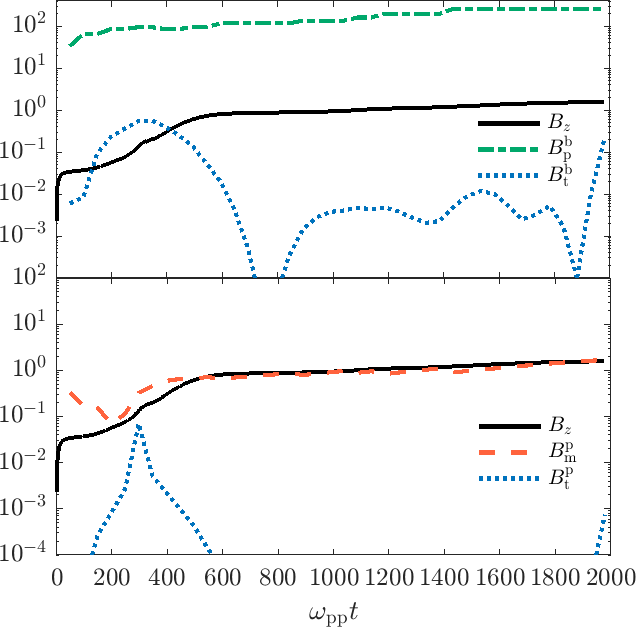} 
          \caption{Temporal evolution of the simulated mean $B$-field strength ($B_z$, black curves) compared to various saturation criteria for reference run (a). Top panel: particle ($B_{\rm p}^{\rm b}$, green dashed-dotted curve) and trapping ($B_{\rm t}^{\rm b}$, blue dotted curve) limits as applied to the beam particles. Bottom panel: spatial magnetization ($B_{\rm m}^{\rm p}$, red dashed curve) and trapping ($B_{\rm t}^{\rm p}$, blue dotted curve) limits as applied to the plasma particles. All curves are in units of $m_e c \omega_{\rm pp}/e$.
}
        \label{fig: B-comp refcase} 
\end{figure}

The growth of the magnetic field during the linear and saturation phases of our reference case (a) can be clearly seen in Fig.~\ref{fig: B-comp refcase} (thick black line). In this figure, and subsequent similar ones, the $B$-field strength is expressed in dimensionless units, $\overline{B}_z = eB_z/ m_e c \omega_{\rm pp} = B_z/\sqrt{4\pi n_p m_e c^2}$. The  expected maximum growth rate is $\Gamma_{\rm w} \simeq 0.02\,\omega_{\rm pp}$ at $k_\perp \simeq 0.6 \,\omega_{\rm pp}/c$, as obtained by solving numerically the dispersion relation of the CFI \cite{Bret_2010b}. This computation also yields $\tilde\chi_{\rm p} \simeq 0.006$ and $\chi_{\rm b} \simeq 0.07$, thus showing that the kinetic limit does apply for both components. For reference, the approximations of Eq.~\eqref{eq:growth_rate_kk} give $\Gamma_{\rm w, k-k} \simeq 0.01 \omega_{\rm pp}$ and $k_\perp \simeq 0.7 \,\omega_{\rm pp} /c$ in that regime. These predictions fairly match the simulations results: the growth rate evaluated using Eq.~\eqref{eq:GR_linear_phase} between $t=200\,\omega_{\rm pp}^{-1}$ and $t=450 \, \omega_{\rm pp}^{-1}$ is $ \Gamma_{\rm w}^{\rm PIC} \simeq 8 \times 10^{-3}\,\omega_{\rm pp}$, while the dominant $k_\perp$ in the Fourier spectrum of $B_z$ at saturation ($t \simeq 500\,\omega_{\rm pp}^{-1}$) is measured to be $k_\perp^{\rm PIC} \simeq 0.8\,\omega_{\rm pp}/c$.
Considering that the spectrum of the instability is rather broad and variable with time, the factor of $\sim 2$ discrepancy between the theoretical and simulation results is not very significant.


The measured value of $\overline{B}_z$ is compared to the saturation limits $B_{\rm t}^{\rm b}$ and $B_{\rm p}^{\rm b}$ in the upper panel of Fig.~\ref{fig: B-comp refcase}, and to $B_{\rm t}^{\rm p}$ and $B_{\rm m}^{\rm p}$ in the lower panel. As explained earlier, we only plot the maximum of the two ``particle limit'' criteria corresponding to either component, since the lower one is not relevant for determining saturation. In the present case, the current density carried by the beam largely dominates that of the plasma because $\vert\beta_{\rm p}\vert\ll1$. We do not plot $B_{\rm m}^{\rm b}$ because it lies far above $B_{\rm t}^{\rm b}$, as expected from the discussion of Sec.~\ref{sec: sym case}.  Recalling that the limits given in Eqs.~\eqref{eq:trapp}, \eqref{eq: max B magn} and \eqref{eq: plimit} are upper limits, saturation is expected to occur once the measured $B$ value exceeds one of the corresponding curves in Fig.~\ref{fig: B-comp refcase}. All limits shown here are computed from the instantaneous quantities measured in the simulation, which explains their evolution in time. A word of caution thus appears necessary regarding $B_{\rm t}$: as it scales with $\Gamma_{\rm w}^2$, which is computed through Eq.~\eqref{eq:GR_linear_phase}, this limit becomes meaningless outside the phase of linear growth of the CFI. In particular, the fact that $\overline{B}_{\rm t}^{\rm b} < \overline{B}_z$ at early times ($t\lesssim 100\,\omega_{\rm pp}^{-1}$), does not mean that saturation has occurred. On the other hand, the fact that $\overline{B}_{\rm t}^{\rm b}$ and $\overline{B}$ cross each other at $t \simeq 400\,\omega_{\rm pp}^{-1}$ is indicative of saturation through trapping.

\begin{figure}
        \includegraphics[width=0.45\textwidth]{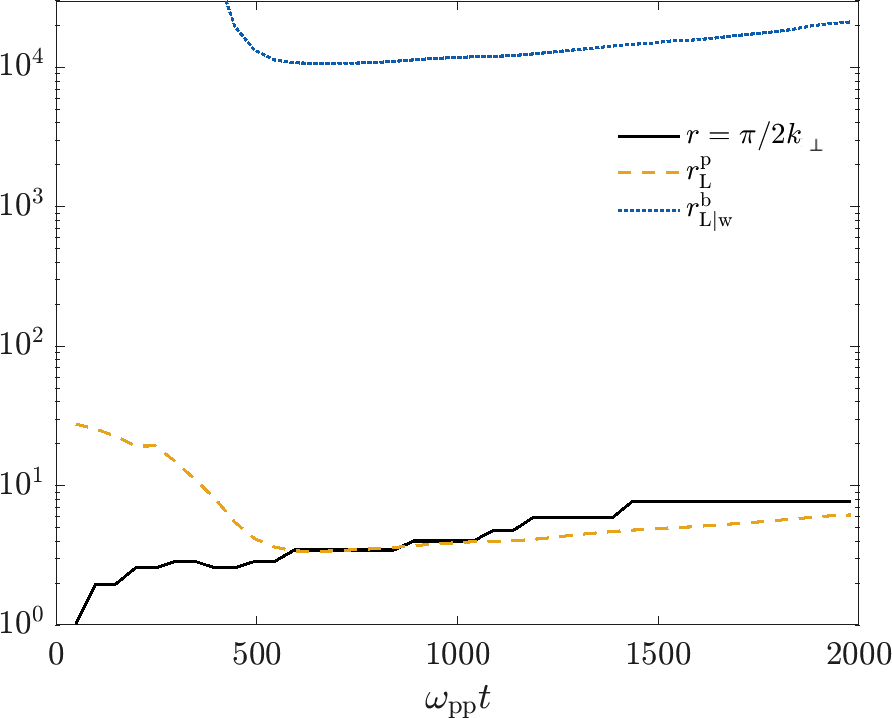} 
        \caption{Comparison of the typical filament size, as extracted from simulation (black line), with the Larmor radius of plasma particles in the simulation frame (red dashed line) and of beam particles in the instantaneous Weibel frame (orange dotted line). Both radii are computed using dynamical quantities extracted from the reference run~(a) as defined in Table~\ref{Tab:sim}.}
        \label{fig: rL refcase}
\end{figure} 

Around $t \simeq 400 \,\omega_{\rm pp}^{-1}$ the magnetic field indeed becomes so strong that the quiver frequency of the beam particles exceeds the growth rate of the instability. Beam particles can then be regarded as transversely \textit{trapped} around the $B$-field nodes (Fig.~\ref{fig: B-comp refcase} top panel). To quantify this, we use the characteristic momentum and Lorentz factor averaged over the Maxwell-J\"uttner distribution  as $\langle \gamma_{\rm b}  \beta_{\rm b \parallel} \rangle \simeq \langle \gamma_{\rm b} \rangle \simeq 4 \gamma_{\rm b}/\mu_{\rm b}$. Combining those values with the theoretical estimates of $\Gamma_{\rm w}, k_\perp$, and the parameters of Table~\ref{Tab:sim}, we derive the trapping limit as $\overline{B}_{\rm t}^{\rm b} \simeq (\Gamma_{\rm w}/\omega_{\rm pp} )^2 (\omega_{\rm pp}/ k_\perp c) \langle \gamma_{\rm b} \rangle/\beta_{\rm b\parallel} \simeq 0.5$, which matches well the observed saturation value $\overline{B}_z \simeq 1$. $\overline{B}_{\rm t}^{\rm b}$ is also close to the estimate from the measured values of $\Gamma_{\rm w}$, $k_\perp$, $\langle \gamma_{\rm b} \rangle$ and $\beta_{\rm b\parallel}$, that is, $\overline{B}_{\rm t}^{\rm b,PIC} \simeq 0.3$. As expected, the particle limit for the beam lies above those values,
$\overline{B}_{\rm p}^{\rm b} \simeq (\pi/2) (n_{\rm b}/n_{\rm p}) (\omega_{\rm pp}/ k_\perp c) \langle \gamma_{\rm b} \beta_{\rm b \parallel} \rangle \simeq 50$,
and the magnetization limit well above, $\overline{B}_{\rm m,r_L}^{\rm b} \simeq ( 2/\pi)(k_\perp c/\omega_{\rm pp}) \langle \gamma_{\rm b} \beta_{\rm b\parallel} \rangle \simeq 1.6 \times 10^3$.   
As anticipated in Sec.~\ref{sec: sym case}, the trapping limit for the beam thus appears to provide the relevant criterion for saturation. Interestingly, $\overline{B}_{\rm t}^{\rm p} \ll \overline{B}_z$ at all times, even during linear growth, indicating that the strong quiver motion of the plasma component does not prevent the CFI from growing, neither does it matter from the point of view of saturation. 

The large value of $\overline B_{\rm m}^{\rm b}$ confirms that magnetic trapping does not act longitudinally, meaning that the Larmor radius of the beam particles remains much larger than the characteristic radius of a filament; see in particular Fig.~\ref{fig: rL refcase} which carries out such a comparison. As already pointed out in Sec.~\ref{sss: magn. limit}, if the drift velocity is relativistic, as is the case for the beam particles, the magnetization limit is determined by the spatial constraint $r_L \lesssim r$.
We recall that the notion of Larmor radius implies a constant $B$-field along with a null electric field, and hence has to be computed in the instantaneous Weibel frame, which departs, given the development of the instability, from the simulation frame. This change of frame is relevant for the beam, which moves relativistically in the simulation frame at $u_b \simeq 19 \simeq \rm const$, while it can be neglected for the background plasma, given that its velocity and the Weibel frame velocity remain sub-relativistic in the simulation frame ($\vert\beta_{\rm p\vert w}\vert \ll 1$). 

\begin{figure}
    \includegraphics[width=0.42\textwidth]{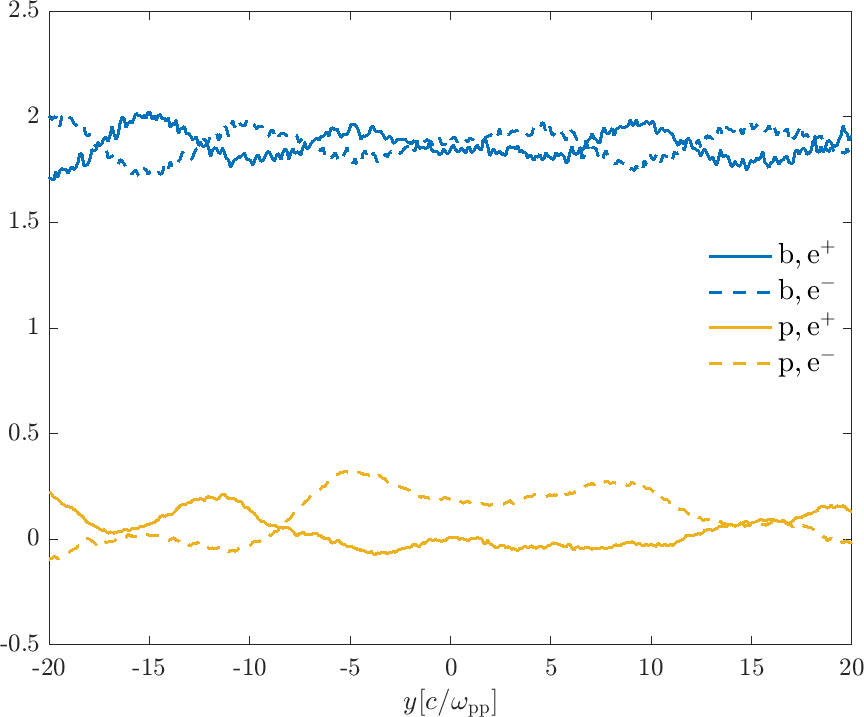}  
    \caption{Transverse profiles of the beam (blue curves) and plasma (yellow curves) current densities, at saturation $t \simeq 400\,\omega_{\rm pp} ^{-1}$ in the reference run (a) listed, and in a limited region of the periodic $y$-domain. For each species, the solid and dashed curves correspond to positrons and electrons, respectively.
    }
    \label{fig: ixiy refcase}
\end{figure}

To better understand why the particle limit does not provide the relevant saturation criterion here, we quantify the contribution of the beam to the total current to this effect. We plot in Fig.~\ref{fig: ixiy refcase} the particle current density ($n \langle \gamma \beta_{\parallel} \rangle c$) of each species in a limited region of the periodic $y$-domain. One can see that the contributions of the beam and the plasma to the electric current density fluctuations are comparable in scale, although the beam dominates the total particle current density, which enters Eq.~\eqref{eq: plimit}. Importantly, charge separation is not complete and the filaments are rather diluted than spatially split. For this reason, the magnetic field associated with the maximum particle limit among the components remains always greater than the simulated value (compare the green and black curves in Fig.~\ref{fig: B-comp refcase}), and therefore does not account for saturation.


Concerning the background plasma, it remains sub-relativistic and relativistically cold in the Weibel frame, hence $\langle \gamma_{\rm p} \rangle \sim \gamma_{\rm p}$ and $\langle u_{\rm p} \rangle \sim \gamma_{\rm p} \beta_{\rm p}$. As previously mentioned, the $B$-field associated with particle trapping inside the filaments is nearly everywhere much smaller than measured in the simulation: $\overline{B}_{\rm t}^{\rm p} \simeq (\Gamma_{\rm w}/\omega_{\rm pp})^2 (\omega_{\rm pp}/k_\perp c) \langle \gamma_{\rm p} \rangle /\beta_{\rm p \parallel} \simeq 0.001$, as can be verified using the above theoretical estimates for $\Gamma_{\rm max}$ and $k_\perp$. As a matter of fact, the bottom panel of Fig.~\ref{fig: B-comp refcase} shows that the background plasma particles are rapidly trapped inside the filaments, , both transversely and longitudinally, since $\overline{B}_{\rm m, r_L}^{\rm p} \simeq (2/\pi) (k_\perp c/\omega_{\rm pp}) \gamma_{\rm p} \beta_{\rm p} \simeq 0.1 < \overline{B}_z$. 
Actually, $\overline{B}_{\rm m, r_L}^{\rm p}$ rapidly approaches $\overline{B}_z$ (at  $\omega_{\rm pp} t \simeq 200)$ and stays remarkably close to it at later times. We do not interpret this as a cause for saturation of the CFI, but rather as a relaxation of the low-inertia background plasma into the strong magnetic fields driven by the large-inertia beam particles. In runs (b), (c) and (d), $\overline B_{\rm m, r_{\rm L}}^{\rm p}$ gets even smaller than $\overline{B}_z$ during linear growth, indicating that plasma particles become magnetically trapped inside the filaments without inhibiting the CFI growth. 


Well beyond saturation, the characteristic filament radius $r$ increases, roughly linearly in time (see Fig.~\ref{fig: rL refcase}), as a consequence of filament coalescence. However, the $B$-field strength as measured in the Weibel frame, that is, $(B_z^2-E_y^2)^{1/2}$, remains approximately constant. The slow evolution of $\overline B$ in the simulation frame results from the slow evolution of the Weibel frame velocity; it is therefore of kinematic origin. Interestingly, Fig.~\ref{fig: rL refcase} shows that the typical Larmor radius of background plasma particles adjusts at all times to the filament radius, $r_{\rm L,p} \sim r$, which implies that those particles gain energy inside the growing filaments. Qualitatively, this process can be related to the chaotic dynamics of particles trapped in an effective potential characterized by the potential four-vector $A_x \sim r B_z$, which tends to bring equipartition between kinetic $\langle p\rangle$ and potential $e A_x/c$ energies, under the approximate conservation of the canonical momentum $\Pi_x=p_x + eA_x/c$. Such equipartition indeed corresponds to $r_{\rm L,p} \sim r$.

\subsection{Scan in parameter space} \label{sec:other PIC}

The parameters of the reference run (a) are such that the beam carries most of the energy density of the system, and its relativistic plasma frequency is the lower among the two. The smaller inertia of the background plasma particles, which remain sub- or mildly relativistic in the Weibel frame, explains why they relax rapidly in the magnetized filamentary structures while the rigid beam current keeps driving the instability. For this reference run, we thus find that the transverse trapping of beam particles provides the relevant criterion for determining the saturation of the CFI. This general picture proves robust (\emph{i.e.}, it applies from runs (a) to (e) in Table~\ref{Tab:sim}) even if the initial parameters are pushed to extreme values, though always in the CFI-dominated regime. 

\begin{figure}
        \includegraphics[width=0.42\textwidth]{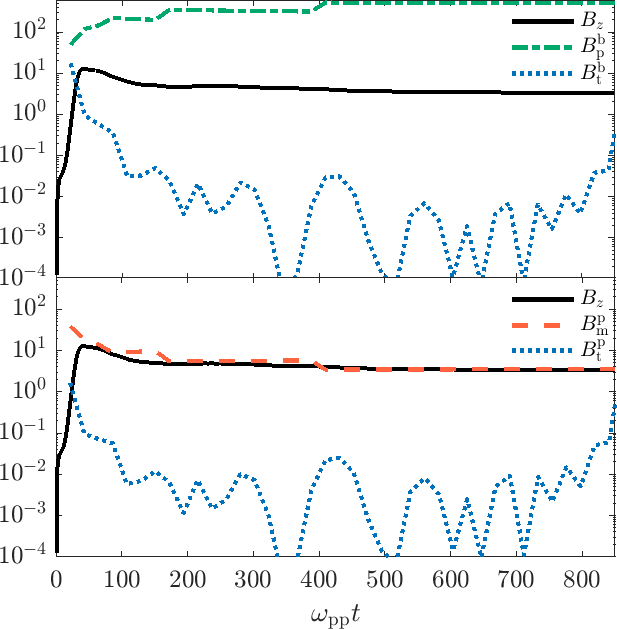}
        \caption{Same as Fig.~\ref{fig: B-comp refcase} for simulation run (e). 
        }
        \label{fig: B-comp Tbddiv10 ecc}
\end{figure}

For instance, Fig.~\ref{fig: B-comp Tbddiv10 ecc} compares the saturation criteria for run (e), in which the initial $\gamma_{\rm b}$ and $T_{\rm p}$ have been multiplied by 10 and $T_{\rm b}$ divided by 10. The instability grows fast, with a measured growth rate $\Gamma_{\rm w}^{\rm PIC} \simeq 0.3\,\omega_{\rm pp}$, $k_\perp^{\rm PIC} \simeq 0.9\,\omega_{\rm pp}/c$, saturating at $t \simeq 30\,\omega_{\rm pp} ^{-1}$. Here as well, transverse magnetic trapping of beam particles appears to control the saturation level, while the Larmor radius of background plasma particles still adapts to the filaments size. The parameters of this simulation, though, are such that Eq.~\eqref{eq:growth_rate_kk} cannot be applied because the plasma is hot, and because it moves at relativistic velocities in the Weibel frame. Solving numerically the dispersion relation of the CFI, we obtain $\Gamma_{\rm w}\simeq 0.3\,\omega_{\rm pp}$ at $k_\perp \simeq 1.2\,\omega_{\rm pp}/c$, which nicely agrees with the PIC values. We then obtain $\overline{B}_{\rm t}^{\rm b} \simeq 37$, a factor of a few above the simulated value $\overline{B}_z \simeq 10$, and slightly below the theoretical particle limit $\overline{B}_{\rm p}^{\rm b} \simeq 44$. The time evolution of these limits, computed with the instantaneous measured values and plotted in the top panel of Fig.~\ref{fig: B-comp Tbddiv10 ecc}, confirms that saturation results from transverse trapping of the beam particles. Moreover, the closeness of the PIC field value and plasma magnetization limit (compare $\overline{B}_z$ and $\overline{B}_{\rm m}^{\rm p}$ in the bottom panel of Fig.~\ref{fig: B-comp Tbddiv10 ecc}) indicates that the plasma particles are fully trapped in the filaments, as before.

Case (f) of Table \ref{Tab:sim}, where $T_{\rm b}$ is reduced by a factor of $30$ while $\gamma_{\rm b}$ and $T_{\rm p}$ are increased by the same amount, provides an exception to that general picture. In this particular configuration, both the beam and the plasma become relativistically hot, leading to comparable initial relativistic plasma frequencies, namely, $\Omega_{\rm pb} \simeq 0.14\,\omega_{\rm pp}$ and $\Omega_{\rm pp} \simeq 0.2\,\omega_{\rm pp}$. One can then hardly discern which plays the role of the beam and which plays the role of the background plasma. What matters for the (transverse or longitudinal) trapping limits, however, is the inertia of the particles. Here, $\langle u_{\rm b}\rangle \simeq 40$ and $\langle u_{\rm p} \rangle \simeq 100$ initially, so that the background plasma particles will be trapped later than the beam particles.

\begin{figure}
        \includegraphics[width=0.42\textwidth]{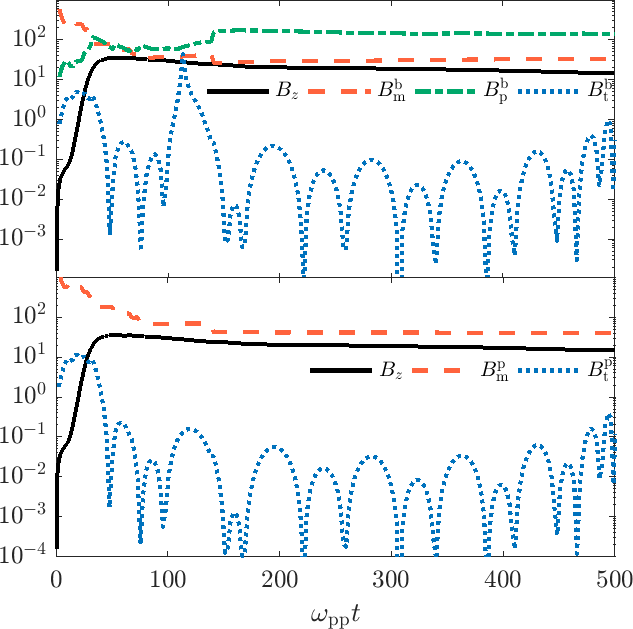}
        \caption{Same as Fig.~\ref{fig: B-comp refcase} for simulation run (f). In the top panel is also plotted the time evolution of the $B$-field associated with the spatial magnetization limit as applied to the beam ($B_{\rm m}^{\rm b}$, red dashed curve).   
        }
        \label{fig: B-comp Tbddiv30 ecc}
\end{figure} 

In detail, we measure $\Gamma_{\rm w}^{\rm PIC} \simeq 0.3\,\omega_{\rm pp}$ and $k_\perp^{\rm PIC} \simeq 3 \omega_{\rm pp}/c$, in fair agreement with the numerical solution to the CFI dispersion relation ($\Gamma_{\rm w} \simeq 0.4\,\omega_{\rm pp}$ at $k_\perp \simeq 1.6\,\omega_{\rm pp}/c$) and which translates into a plasma trapping limit, $\overline{B}_{\rm t}^{\rm p} \simeq 38$, exceeding the beam trapping limit, $\overline{B}_{\rm t}^{\rm b} \simeq 15$. Moreover, since the plasma now carries a larger current density than the beam, it gives a greater particle limit: $\overline{B}_{\rm p}^{\rm p} \simeq 180$ vs. $\overline{B}_{\rm p}^{\rm b} \simeq 38$. Those limits have been evaluated using the simulation parameters; they qualitatively match (yet overestimate by a factor of a few) the values obtained using the instantaneous simulation parameters (as plotted in Fig.~\ref{fig: B-comp Tbddiv30 ecc}). We therefore expect saturation to be determined by transverse plasma trapping as confirmed by Fig.~\ref{fig: B-comp Tbddiv30 ecc}.

In summary, we observe that the CFI growth rate is set by the species with the lower (relativistic) plasma frequency, while the saturation level is determined by that component with the larger inertia per particle, according to the transverse trapping criterion. The expected overall $B$-field amplitude at saturation can thus be approximated as
\begin{equation}
  \frac{B_{\rm sat.}}{\sqrt{4\pi n_{\rm p}m c^2}} \simeq \left( \frac{\Gamma_{\rm w}}{\omega_{\rm pp}} \right)^2 \frac{\omega_{\rm pp}}{k_\perp c} \,,{\rm max} \left( \langle \gamma_{\rm b} \rangle ,\langle \gamma_{\rm p} \rangle \right) \,.
\end{equation}

In the forthcoming section, we extend this analysis to electron-ion compositions.

\section{The electron--ion  case}\label{sec:epCFI}

The presence of ions introduces a new scale in the problem, associated with the hierarchy $m_i/m_e$ ($m_i$ ion mass). If both ions and electrons are cold, the ratio of ion to electron plasma frequencies scales in proportion to $\sqrt{m_e/m_i}$.  If the electrons are heated to such a degree that their effective inertia becomes similar to that of the ion species, then the above hierarchy disappears: both species share a similar relativistic plasma frequency, and hence the electron-ion component effectively behaves as a pair plasma. Thus, one may expect to obtain results similar to those for the pair systems examined in the previous section.

In the particular context of relativistic shock physics, it is known that electrons are efficiently heated up to near equipartition in ultrarelativistic, weakly magnetized conditions (see \emph{e.g.}~\cite{Sironi_2015, 2020Galax...8...33V} and references therein). By contrast, in the mildly relativistic and magnetized regime, electron heating appears to be weak, implying that some hierarchy between the response of electrons and ions remains preserved. Both situations will be addressed in the following.
In order to be able to capture the physics of the instability for both electron and ion species,
with a sufficient number of macro-particles per cell and spatial extent,
we will adopt an ion-to-electron mass ratio $m_i/m_e = 100$.


\begin{table}
    \centering
    \begin{tabular}{ |p{0.6cm}|p{0.6cm}|p{1.2cm}|p{1.cm}|p{1.cm}|p{1.cm}|p{1.cm}|p{0.6cm}|} 
    \hline
 Run                                  	              &   $\gamma_{\rm b}$ & $\gamma_{\rm p}$ & $\frac{T_{\rm be}}{\gamma_{\rm b}}$  & $\frac{T_{\rm bi}}{\gamma_{\rm b}}$ & $\frac{T_{\rm pe}}{\gamma_{\rm p}}$ & $\frac{T_{\rm pi}}{\gamma_{\rm p}}$  & $\frac{\gamma_{\rm b} n_{\rm b}}{\gamma_{\rm p} n_{\rm p}}$ \\ 
    \hline
 (i1)                       	               & $18.9$            &  $1.011$      &            $2.4$   & $0.024$             &           $0.20$                &  $0.0020$ &   $1.9$ \\ 
 (i2)      & $6.7$         &  $1.00075$    &            $4.5$            &           $0.15$      &           $0.2$                          &           $0.002$                &     $1.3$     \\    \hline
    \end{tabular}
    \caption{Parameters of the electron-ion simulations, once transformed to the Weibel frame. Run (i1) is analogous to run (a) in which the positrons have been replaced with ions. Run (i2) treats a mildly relativistic regime. 
    Electron and ion temperatures are given in units of $m_e c^2/k_{\rm B}$ and $m_i c^2/k_{\rm B}$, respectively.
    }
    \label{Tab:simIons}
\end{table}

\subsection{Ultrarelativistic regime} \label{sec:eprel}

Let us first examine the saturation criteria for case (i1) described in Table~\ref{Tab:simIons}. The parameters of this run are obtained from run (a) by replacing the positrons with ions of charge $+e$ and mass $m_i=100\,m_e$. The beam electrons and ions then have a comparable inertia: $\langle p_{\rm be} \rangle/m_e c \simeq 4 \gamma_{\rm b}T_{\rm be} \simeq 3430$ and  $\langle p_{\rm bi} \rangle/m_e c \simeq \gamma_{\rm b} \beta_{\rm b} (m_i/m_e) K_3(\mu_{\rm bi})/K_2(\mu_{\rm bi}) \simeq  4520$ ($K_n$ is the modified Bessel function of the $n$th kind). 
Note that the beam ions have a proper temperature $T_{\rm bi}/m_i c^2 \simeq 0.45$, so that they cannot be considered as fully relativistic.


 
Accounting for the ion mass modifies the trapping and magnetization limits as 
\begin{equation}
    B_{\rm t}^{\rm i} = \frac{\Gamma_{\rm w}^2}{k_\perp} \frac{\langle\gamma\rangle m_i}{\beta_\parallel e}
\end{equation}
and 
\begin{equation}
    B_{\mathrm{m},\,r_\mathrm{L}}^{\rm i}=  \frac{2}{\pi} k_\perp \langle \gamma \beta \rangle \frac{m_i c^2}{e} \,.
\end{equation}

\begin{figure}
        \includegraphics[width=0.42\textwidth]{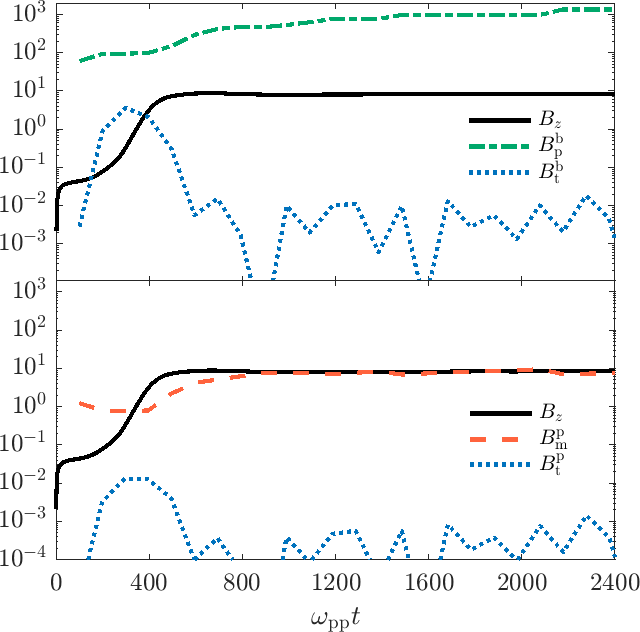}
        \caption{Temporal evolution of the simulated mean $B$-field strength (black curves) compared to various saturation criteria for run (i1) defined in Table~\ref{Tab:simIons}. Top panel: particle (green dashed-dotted curve) and trapping (blue dotted curve) limits as applied to the beam ions. Bottom panel: spatial magnetization (red dashed curve) and trapping (blue dotted curve) limits as applied to the plasma ions. All curves are in units of $m_e c \omega_{\rm pp}/e$.}
        \label{fig: B-comp ref. case e-Ions}
\end{figure}

The time evolution of the simulated mean $B$-field is plotted in Fig.~\ref{fig: B-comp ref. case e-Ions}. Unlike previous studies (\emph{e.g.} \cite{Ruyer_2015a}), the system does not  experience an early phase governed by electrons, in which the CFI grows faster, before moving to a regime ruled by the slower ion-driven CFI. We ascribe this behavior to the similar inertia of the beam ions and electrons.
Solving the CFI dispersion relation in the presence of ions yields a maximum growth rate $\Gamma_{\rm w} \simeq 0.025\, \omega_{\rm pp} $ for a wavenumber $k_\perp \simeq 0.5\,\omega_{\rm pp}/c$ (as before, $\omega_{\rm pp}$ denotes the \emph{electron} plasma frequency of the background plasma). These values are very close to the simulation values, namely, $\Gamma_{\rm w}^{\rm PIC} \simeq 0.024\,\omega_{\rm pp}$ (as obtained by exponentially fitting $B_z(t)$ over $300 < \omega_{\rm pp} t < 500$) and $k_\perp^{\rm PIC} \simeq 0.4\,\omega_{\rm pp}/c$ (as measured from the spatial Fourier spectrum of $B_z$).

As in run (a), the CFI saturates through transverse trapping of the beam particles (electrons and ions). This is consistent with the fact that the theoretical trapping limit, $\overline{B}_{\rm t}^{\rm b} \simeq 6$, is much smaller than the particle limit, $\overline{B}_{\rm p}^{\rm b} \simeq 75$, both limits being computed for the beam ions and using the initial simulation parameters. This estimate of $\overline{B}_{\rm t}^{\rm b}$ matches well that evaluated at saturation time ($t\simeq 400\,\omega_{\rm pp}^{-1}$) instantaneous simulation parameters (see top panel of Fig.~\ref{fig: B-comp ref. case e-Ions}). At later times, again similarly to run (a), the background plasma particles turn fully magnetized, with their typical Larmor radius adjusting to the mean filament size (bottom panel of Fig.~\ref{fig: B-comp ref. case e-Ions}). A notable difference with run (a), however, is that the mean $B$-field strength here remains quasi-constant following saturation (up to the final simulation time, $t=2400\,\omega_{\rm pp}^{-1}$), rather than slowly increasing as in Fig.~\ref{fig: B-comp refcase}.


In short, in this asymmetric, relativistic electron-ion simulation, in which both species share a similar inertia, we recover the general picture of the previous section. Accordingly, the CFI saturation is determined by the trapping limit as applied to the species with the largest inertia.

\subsection{Mildly relativistic regime} \label{sec:epnrel}

\begin{figure}
        \includegraphics[width=0.42\textwidth]{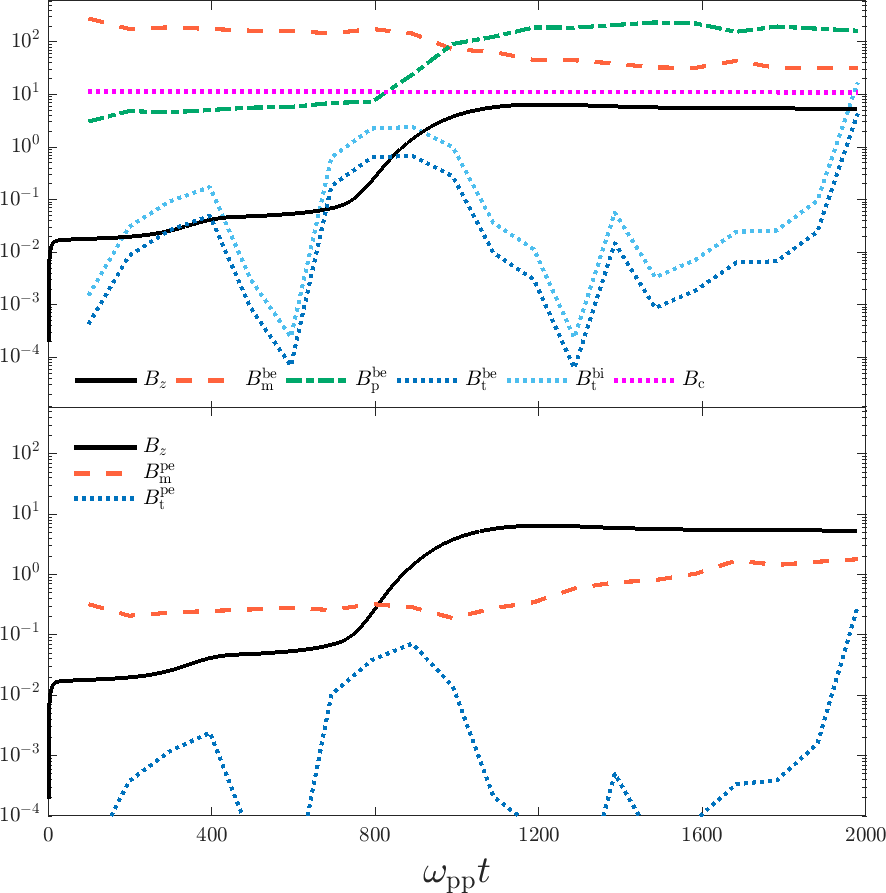}
        \caption{Temporal evolution of the simulated mean $B$-field strength ($B_z$, black curves) compared to various saturation criteria for run (i2) defined in Table~\ref{Tab:simIons}. Top panel: comparison of the spatial magnetization ($B_{\rm m}^{be}$, red dashed curve), particle ($B_{\rm p}^{be}$, green dashed-dotted curve), trapping ($B_{\rm t}^{\rm be}$, blue dotted curve) limits as applied to the beam electrons, plus the trapping limit applied to beam ions ($B_{\rm t}^{\rm bi}$, light-blue dotted curve). Also plotted is the saturated $B$-field from Eq.~\eqref{eq: Bsat Peterson} ($B_{\rm c}$, magenta dotted line).
        Bottom panel: spatial magnetization ($B_{\rm m}^{\rm pe}$, red dashed curve) and trapping ($B_{\rm t}^{\rm pe}$, blue dotted curve) limits as applied to the plasma electrons. All curves are in units of $m_e c \omega_{\rm pp}/e$.
        }
        \label{fig: B-comp i2}
\end{figure}

We now address the case of two electron-ion plasmas counterstreaming with a moderate Lorentz factor ($\sim 3$) in a reference frame. These two plasma flows mainly differ in their temperatures: the beam's electron and ion populations are much hotter than their plasma counterparts, and for each (beam or plasma) component, the electrons are also much hotter than the ions. In particular, the difference in temperature between the beam ions and electrons is justified by the fact that, according to kinetic simulations, the shock-reflected ions have a temperature at least three times larger than their electronic counterpart in the downstream frame (see \cite{Caprioli_2014, Crumley_2019, Ligorini_2021a, Ligorini_2021b} and references therein).
The initial parameters for this run (i2), as expressed in the corresponding Weibel frame, are summarized in Table~\ref{Tab:simIons}.

In this configuration, one has $\langle p_{\rm be}\rangle \simeq 800\,m_e c$, while $\langle p_{\rm bi}\rangle \simeq 3000\,m_e c$. A hierarchy therefore persists between the beam electrons and ions, leading to a somewhat different picture for the evolution of the instability and its saturation level.

Figure~\ref{fig: B-comp i2} shows that after a transient early phase ruled by oblique modes, the CFI sets in at $t \simeq 300\,\omega_{\rm pp}^{-1}$ and rapidly saturates at $t \simeq 400\,\omega_{\rm pp}^{-1}$ with a measured growth rate $\Gamma_{\rm w}^{\rm PIC} \simeq 5 \times 10^{-3}\,\omega_{\rm pp}$ and a dominant wave number $k_\perp^{\rm PIC} \simeq 0.35\,\omega_{\rm pp}/c$. During this short period, the $B$-field grows only by a factor of a few, likely because the transverse trapping limit for beam electrons is already partially fulfilled, see top panel of Fig.~\ref{fig: B-comp i2}. This figure also suggests that the trapping of beam ions contributes to the instability saturation, as would be expected from their larger inertia.

\begin{figure}
        \includegraphics[width=0.49\textwidth]{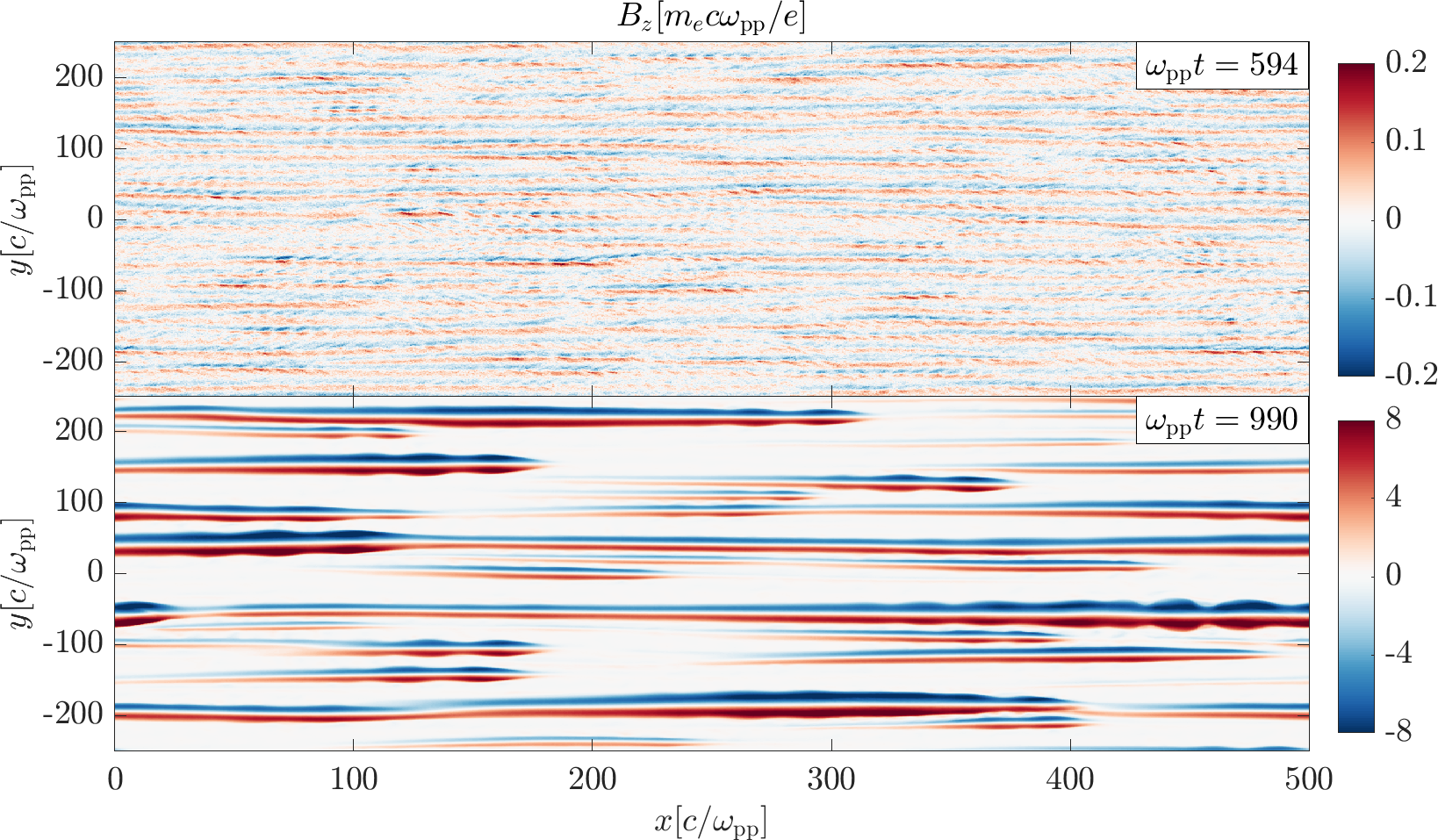}
        \caption{Out-of-plane magnetic field ($B_z$) generated by the counterstreaming of mildly relativistic electron-ion flows (i2). The magnetic field is plotted at two different times: in the early ($\omega_{\rm pp}t =594$, top) and  late ($\omega_{\rm pp}t =990$, bottom) phases of the CFI when cavities have started to form.
        }
        \label{fig: 2D Bz mid-rel}
\end{figure}

\begin{figure}
        \includegraphics[width=0.48\textwidth]{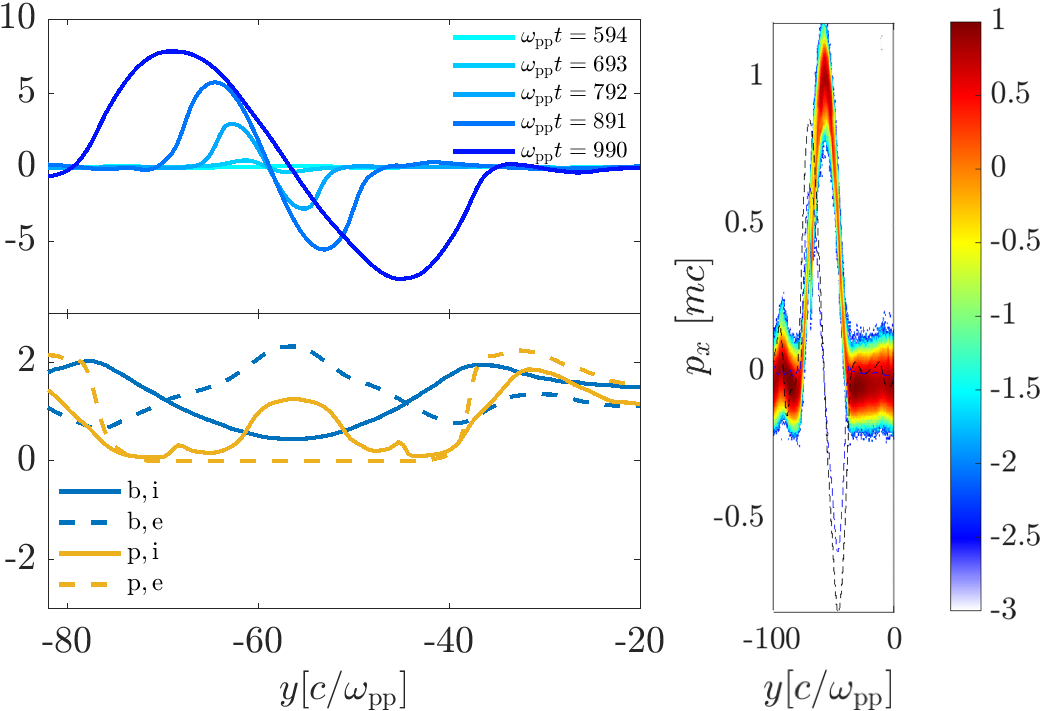}
        \caption{Top left panel: magnetic field profiles along the transverse direction ($y$) and at successive times, as indicated, for simulation run (i2). The figure reveals the growth of the magnetic field as the cavity expands. Bottom left panel: transverse profiles of the number density of the beam and plasma components at the onset of saturation, $\omega_{\rm pp} t \simeq 1000$. Right panel: $(y,p_x)$ phase space of the plasma ions at the same time.}
        \label{fig: n-species mid-rel}
\end{figure}

At later times ($t\gtrsim 700\omega_{\rm pp}^{-1}$), a secondary instability develops, leading the mean $B$-field strength to rise by almost two orders of magnitude. As shown in Fig.~\ref{fig: 2D Bz mid-rel},  this instability generates isolated, large-scale magnetic filamentary structures, which are essentially filled with beam electrons and plasma ions, and devoid of beam ions and plasma electrons.
Those structures, or ``cavities'', have been observed in previous electron-ion simulations  ~\cite{Ruyer_2015b, Naseri+_2018} and studied recently in greater detail in Ref.~\cite{Peterson+_2021}.
Although the latter paper considered a simpler setting consisting of an electron beam-plasma system embedded in an ion background, the picture that it sketches can be readily extended to the present problem.
Specifically, the cavities
are driven by the beam electrons, which are initially overdense relative to the plasma ($\gamma_{\rm be} n_{\rm be}/ \gamma_{\rm pe} n_{\rm pe} \simeq 1.3$ at $t=0$). As a cavity expands due to the magnetic pressure exerted by the beam electron current, more beam electrons join and add their contribution to the current inside the cavity, thus feeding back positively on the magnetic field. Meanwhile, the beam ions are expelled from the cavity by the growing field, just as the plasma electrons. The background ions accumulate in the cavity, mainly (initially) as a result of the confining force exerted by the $E_y$ electric field component. This scenario is illustrated in Fig.~\ref{fig: n-species mid-rel}, in the case of the cavity formed at $(x,y) \simeq (450,-60)\,c/\omega_{\rm pp}$ in the bottom panel of Fig.~\ref{fig: 2D Bz mid-rel}. The top left panel depicts the time evolution of the $B$-field profile across the cavity, while the bottom left panel plots the density profiles of the various populations of the system.

Interestingly, this secondary instability is essentially driven by one species, here the beam electrons, and it leads to a sharp contrast between the beam electron density inside and outside the cavity. Thus, it is not surprising that the particle limit, as evaluated for the beam electrons, nicely follows the evolution of the magnetic field during this nonlinear phase\footnote{In Fig.~\ref{fig: B-comp i2} there is an offset of about an order of magnitude between the measured value $B_z$ (black solid curve) and the theoretical limit corresponding to \eqref{eq: plimit} (green dashed-dotted curve). This offset is related to the overall geometry, in particular the fact that the structures are not space-filling while the averages are taken over the simulation box. It is clear, however, that inside a cavity, the magnetic field is mostly carried by the beam electrons.}, yet this does not cause the instability to saturate.

We also note that a key factor for this secondary instability is a clear hierarchy between the beam ions and the beam electrons. Were they of equal inertia, these two species would react similarly in adjacent filaments, leading to the growth of all filaments as in the standard CFI. A comparison of this simulation with the previous one (i1) suggests that in order for the instability to develop, the beam electrons and ions should differ in their inertia by at least a factor of a few. 

This instability causes the magnetic field to grow rapidly until saturation is reached at $t\simeq 1100\,\omega_{\rm pp} ^{-1}$.
While in \cite{Peterson+_2021}, the magnetic pressure pushes a ``wall'' composed of background ions initially at rest, in the present case it evacuates the beam ions, which are relativistic. We can thus adapt the calculation of the instability growth rate made in that study to our conditions by taking into account the inertia of the beam ions, as follows.

Assuming that the $B$-field inside the cavity is mainly generated by the beam electrons, the magnetic pressure acting on this wall can be expressed as 
\begin{equation}
    \frac{B_z^2}{8 \pi} = \frac{(4 \pi e  n_{\rm be} \gamma_{\rm be} \beta_{\rm be} r_{\rm c})^2}{8 \pi} \,.
\end{equation}
The momentum per unit area of the wall is mainly carried by the expelled beam ions, and so can be estimated as $\gamma_{\rm bi} n_{\rm bi} \langle p_{\rm bi} \rangle r_{\rm c}(t)$, where $r_{\rm c}(t)$ is the instantaneous cavity radius. Momentum balance in the transverse ($y$) direction then leads to
\begin{equation}
  \frac{d}{dt}\left( \gamma_{\rm bi} n_{\rm bi} \langle p_{\rm bi} \rangle r_{\rm c} \frac{dr_{\rm c}}{dt}\right)  = 
  2\pi (e n_{\rm be} \gamma_{\rm be }\beta_{\rm be} r_{\rm c})^2 \,,
\end{equation}
The solution to this equation grows as $r_{\rm c} \propto e^{\Gamma_{\rm c} t}$, where the growth rate is given by
\begin{equation}
    \Gamma_{\rm c} = \frac{\Omega_{\rm pbi}}{2} \,.
    \label{eq: grRate mildly rel}
\end{equation}
In the present case, $\Omega_{\rm pbi} \simeq 0.020\,\omega_{\rm pp}$, which is in fair agreement with the growth rate $\Gamma^{\rm PIC} \simeq 0.015 \,\omega_{\rm pp}$ measured in the simulation over the interval $750 \lesssim \omega_{\rm pp} t \lesssim 960$.

According to \cite{Peterson+_2021}, saturation is reached once the background plasma ions are accelerated by the inductive electric field ($E_x$) to a point where they become relativistic ($p_{i,x} \simeq m_i c$) and neutralize the electron beam current. The right panel of Fig.~\ref{fig: n-species mid-rel}, which displays the $(y,p_x)$ phase space of the plasma ions at the onset of saturation ($\omega_{\rm pp}t = 990 $), confirms that they have indeed attained relativistic momenta by that time inside the cavity. Adapting again the calculations in Ref.~\cite{Peterson+_2021}, the radius of the cavity at saturation can be expressed as 
\begin{equation}
    r_{\rm c, sat} = \frac{c}{ \gamma_{\rm b}^{1/2} \omega_{\rm pbi}} = \frac{c}{\langle \gamma_{\rm bi}\rangle^{1/2} \Omega_{\rm pbi}} \,,
\end{equation}
recalling that $\omega_{\rm pbi} = \sqrt{4 \pi n_{\rm bi}e^2/m_i}$. This gives $r_{\rm c, sat} \simeq 10\,c/\omega_{\rm pp}$, which agrees relatively well with the size of the structures seen in Figs.~\ref{fig: 2D Bz mid-rel} and \ref{fig: n-species mid-rel}.

The corresponding saturated value of the magnetic field is given by 
\begin{equation}
    \overline{B}_{\rm c} \simeq \gamma_{\rm be}^{1/2}\left(\frac{m_i}{m_e}\right)^{1/2} \frac{\omega_{\rm pbe}}{\omega_{\rm pp}}
    \label{eq: Bsat Peterson}
\end{equation}
in normalized units. One obtains $\overline{B}_{\rm c} \simeq 11$ in correct agreement with the observed value $\overline{B}_z \simeq 6$ (see top panel of Fig.~\ref{fig: B-comp i2}). Note that in Ref.~\cite{Peterson+_2021} an extra factor of $1/2$ was added in the estimation of the saturated field, which is not included here.



\section{Late--time evolution of the beam-plasma system}
\label{subsec: late time ev}

We conclude by investigating briefly the late time evolution of the beam-plasma system after the saturation of the magnetic field. It is worth noting that in this final stage, both the beam and plasma components are expected to relax to isotropy in the turbulence frame. This can be seen as a transition from the two-stream collisionless system to a long-term hydrodynamical system in which everything has been effectively mixed. In this respect, if we assume that the beam and plasma have relaxed to the same final velocity but with different temperatures, the conservation of energy and momentum implies: 
\begin{align}
    \gamma_{\rm bi}^2 w_{\rm bi} -p_{\rm bi} + \gamma_{\rm pi}^2 w_{\rm pi} -p_{\rm pi} & = \gamma_{\rm f}^2(w_{\rm bf}+w_{\rm pf}) \\ \nonumber 
    &- p_{\rm bf} - p_{\rm pf} \,, \\
    \gamma_{\rm bi}^2\beta_{\rm bi}w_{\rm bi} + \gamma_{\rm pi}^2\beta_{\rm pi} w_{\rm pi} &= \gamma_{\rm f}^2\beta_{\rm f}(w_{\rm bf}+w_{\rm pf}) \,,
\end{align}
where the subscripts $_{\rm i}$ and $_{\rm f}$ here refer, respectively, to the initial and final states of the beam ($_{\rm b}$) and plasma ($_{\rm p}$) components. As before, $w$ denotes the enthalpy density and $p$ the pressure.
Note that we have neglected the contribution of magnetic turbulence in the final state, as it is expected to be subdominant. 

In the case where the final states of the beam and plasma are relativistically hot, and therefore share the same adiabatic index, $\widehat{\Gamma}_{\rm f} = w_{\rm f}/(w_{\rm f}-p_ {\rm f})$ ($w_{\rm f}$ and $p_{\rm f}$ are the total final enthalpy density and pressure), the final velocity $\beta_{\rm f}$ satisfies
\begin{equation}
    \frac{\gamma_{\rm bi}^2 w_{\rm bi} -p_{\rm bi} + \gamma_{\rm pi}^2 w_{\rm pi} -p_{\rm pi}}{\gamma_{\rm bi}^2\beta_{\rm bi}w_{\rm bi} + \gamma_{\rm pi}^2\beta_{\rm pi} w_{\rm pi}} =
    \frac{\kappa_{\rm f} -1+\beta_{\rm f}^2 }{\kappa_{\rm f}  \beta_{\rm f}} \, ,
\end{equation}
where $\kappa_{\rm f} \equiv \widehat{\Gamma}_{\rm f}/(\widehat{\Gamma}_{\rm f}-1 )$.

Consider for instance the case, exemplified by run (a) of Table~\ref{Tab:sim}, of an initially sub-relativistic ($\beta_{\rm p,i} \sim 0$) and cold ($p_{\rm p,i} \sim 0$) plasma interacting with a relativistically hot beam which carries most of the energy ({\it i.e.}  $\gamma_{\rm b,i}^2\,p_{\rm b,i} \gg w_{\rm p,i}$). We then have $\kappa_{\rm f} \simeq 4$ (as in the initial state), so that  
\begin{align}
    \beta_{\rm f}&\,  \simeq\, \beta_{\rm b}\left(1 - \frac{1}{2}\, \frac{w_{\rm pi}}{ \gamma_{\rm bi}^2 p_{\rm bi}}\right),\nonumber\\
    \gamma_{\rm f}&\,  \simeq\, \gamma_{\rm b}\left(1 - \, \frac{w_{\rm pi}}{p_{\rm bi}}\right)\,.
    \label{eq: gam hydro}
\end{align}
The second equation further assumes $w_{\rm pi}\ll p_{\rm bi}$. The Lorentz factors are evaluated in the simulation frame of the two-stream system. 

\begin{figure}
        \includegraphics[width=0.45\textwidth]{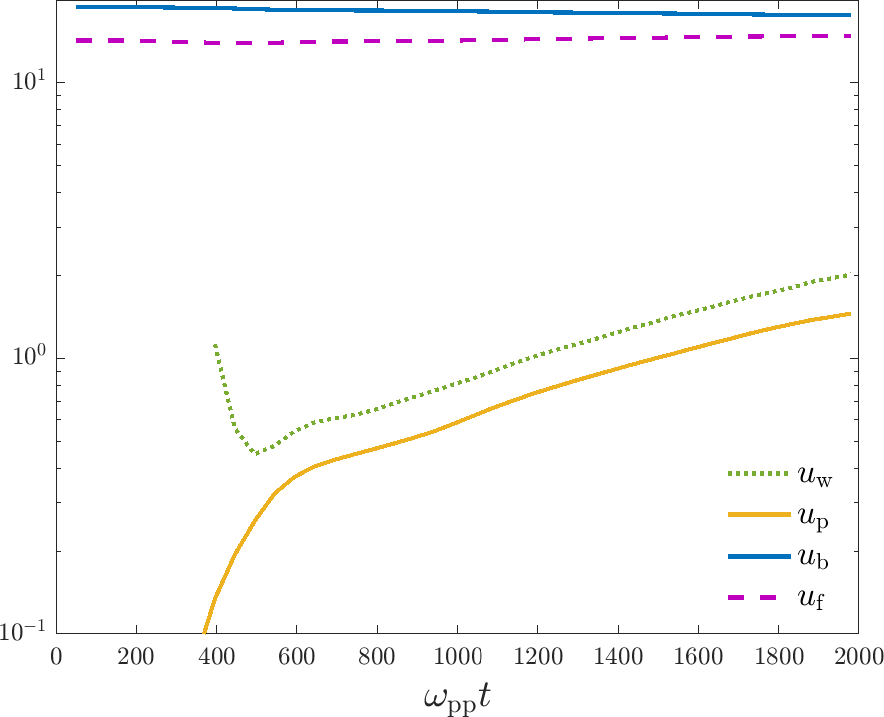}
        \caption{Time evolution of various four-velocities as extracted from run (a) defined in Table~\ref{Tab:sim}. Light-green dotted curve: four-velocity of the Weibel frame. Yellow curve: four-velocity of the plasma. Blue curve: four-velocity of the beam. Magenta dashed line: four-velocity of the relaxed plasma and beam as given by Eq.~\eqref{eq: gam hydro}.
        }
        \label{fig: 4velocity refcase}
\end{figure} 

The above indicates that the asymptotic velocity of the relaxed components should be close to the initial beam velocity.  This behavior is illustrated in Fig.~\ref{fig: 4velocity refcase}, which shows the time evolution of various four-velocities as extracted from our reference run (a). The beam four-velocity $u_{\rm b}$ (blue) indeed approaches from above the predicted asymptotic four-velocity, $u_{\rm f} = \beta_{\rm f} \gamma_{\rm f}$ (magenta dashed line). Conversely, the plasma four-velocity $u_{\rm p}$ is seen to increase steadily toward $u_{\rm f}$. Also overlaid is the instantaneous four-velocity of the Weibel frame (green dotted line), computed from the simulation data as $u_{\rm w}=\gamma_{\rm w} \beta_{\rm w}$ with $\beta_{\rm w} = (\langle E_y^2 \rangle/\langle B_z^2\rangle)^{1/2}$ (the average is taken over the simulation domain). Note that this quantity is not defined at early times because of the dominance of oblique modes characterized by $\langle E_y^2 \rangle/\langle B_z^2 \rangle > 1$. The four-velocity of the Weibel frame tracks that of the background plasma quite well. 
We note that the convergence to the hydrodynamical regime is not attained over the time scale of the simulation. As a matter of fact, we expect the convergence to proceed at a slower rate as time increases. This is because relaxation takes place in the Weibel frame, hence time dilation effects associated with the relativistic velocity of the Weibel frame relative to the simulation frame will slow down the apparent relaxation rate. 

\section{Conclusions} \label{sec:conclusion}

In this paper, we have investigated the saturation of the current filamentation instability, or Weibel instability, in an asymmetric configuration, meaning in the case in which the counterstreaming plasmas differ in terms of velocity, temperature and density. This configuration is notably representative of the precursor region of electron-positron or electron-ion shocks, although the implications of our results are not restricted to such systems. Our study relies on large-scale periodic PIC simulations of counterstreaming flows composed of a hot dilute population representing the beam (\emph{e.g.} the particles reflected at the shock front) and a relatively cold plasma \emph{e.g.} the background plasma that is incoming toward the shock). The parameters of our fiducial run have been directly borrowed from a large-scale relativistic shock simulation at a position deep in the precursor; the parameters of subsequent runs have then been varied in an \emph{ad hoc} manner to explore different possible settings. We have discussed several theoretically motivated criteria for saturation and compared them to the simulation results.

The asymmetric counterstreaming configuration departs from its symmetric counterpart in two important ways: (1) there exists an ambiguity as to whether a given criterion should be applied to the beam, or to the plasma component; (2) there exists a preferred reference frame, dubbed here the ``Weibel frame''~\cite{Pelletier+_2019}, in which the instability is purely magnetic; this reference frame does not \emph{a priori} coincide with that in which the total momentum flux vanishes, as happens for the symmetric configuration. Here, we pay particular attention to that latter point. We have set up our simulations such that for each set of parameters characterizing the plasma flows, the simulation frame initially coincides with the Weibel frame.

We have then compared different mechanisms as possible sources of saturation of the magnetic field associated with the instability: magnetic trapping, particle limit, Alfv\'en limit.  Our general conclusion is that, for pair plasmas, the saturation level is determined by the criterion of magnetic trapping as applied to the (beam or plasma) component that carries the larger inertia of the two: the growth rate is found to diminish strongly once the quiver frequency of that component becomes comparable with, or larger than the instability growth rate. For all studied cases, our theoretical estimates of the instability properties, such as the maximum growth rate and associated wave number, are consistent with those extracted from the simulations. Consequently, it is possible to obtain reasonable analytical approximations for the strength of the magnetic field at saturation. Furthermore, we find that the particle limit is never fulfilled, all the more so when the component of larger inertia is relativistically hot, as its temperature then prevents its charged species from being fully segregated in separate filaments. We have observed that the component of smaller inertia becomes rapidly trapped inside the filaments, in some cases even during the linear phase of the CFI. At late times, the Larmor radius of those particles closely follows the characteristic filament radius and thus grows in time through coalescence. Asymptotically, the system tends to a final state where the two fluids are effectively mixed, drifting at the same mean velocity. However, due to relativistic time dilation effects, this ultimate regime could not be accessed from our simulations.

We have also investigated the case of asymmetric electron-ion systems with a mass ratio $m_i/m_e=100$. As long as there is not a clear hierarchy in inertia between the electron and ions species of a given (beam or plasma) component at the beginning of the simulation, the development of the instability and the saturation proceed much as in the case of a pair plasma. The picture and saturation criterion discussed above thus remain applicable. However, if the electron and ion inertia differ by a factor of a few or more, a different instability eventually supersedes the CFI. It leads to the formation of cavities in which the beam electrons and background plasma ions accumulate and drive magnetic field growth, while the beam ions are pushed outwards along with the plasma electrons. This mechanism comes to an end when the plasma ions inside the cavities, accelerated by the inductive electric field, become capable of neutralizing the electron beam current, as discussed recently in Ref.~\cite{Peterson+_2021}.

\acknowledgements

This work was supported by the ANR (UnRIP project, Grant No.~ANR-20-CE30-0030). We acknowledge GENCI-TGCC for granting us access to the supercomputer IRENE under Grants No.~2019-A0050407666, 2020-A0080411422 and 2021-A0080411422 to run PIC simulations. We acknowledge financial support from Centre National d'\'Etudes Spatiales (CNES). This work has also been supported in part by the Sorbonne Universit\'e DIWINE Emergence-2019 program.


\appendix

\bibliographystyle{apsrev4-1}
\bibliography{mybib}
\end{document}